\documentclass{JINST}

\usepackage{graphicx}
\graphicspath{{figs}}

\usepackage{xspace}
\usepackage{url}

\usepackage{amsmath}

\usepackage{ifthen}

\newboolean{pdflatex}
\setboolean{pdflatex}{false}

\newboolean{uprightparticles}
\setboolean{uprightparticles}{false}
\usepackage{amssymb}
\usepackage{amsfonts}
\usepackage{upgreek}

\def\lhcb {\mbox{LHCb}\xspace}
\def\ux85 {\mbox{UX85}\xspace}

\ifthenelse{\boolean{uprightparticles}}%
{

 \def\PDelta      {\ensuremath{\Delta}\xspace}                 
 \def\PXi      {\ensuremath{\Xi}\xspace}                 
 \def\PLambda      {\ensuremath{\Lambda}\xspace}                 
 \def\PSigma      {\ensuremath{\Sigma}\xspace}                 
 \def\POmega      {\ensuremath{\Omega}\xspace}                 
 \def\PUpsilon      {\ensuremath{\Upsilon}\xspace}                 
 
 \def\PB      {\ensuremath{\mathrm{B}}\xspace}                 
                  
 \def\PD      {\ensuremath{\mathrm{D}}\xspace}

 \def\PK      {\ensuremath{\mathrm{K}}\xspace}

 \def\Pb      {\ensuremath{\mathrm{b}}\xspace}

 \def\Pi      {\ensuremath{\mathrm{i}}\xspace}

}
{

 \mathchardef\PDelta="7101
 \mathchardef\PXi="7104
 \mathchardef\PLambda="7103
 \mathchardef\PSigma="7106
 \mathchardef\POmega="710A
 \mathchardef\PUpsilon="7107
                  
 \def\PB      {\ensuremath{B}\xspace}                 
                  
 \def\PD      {\ensuremath{D}\xspace}

 \def\PK      {\ensuremath{K}\xspace}

 \def\Pb      {\ensuremath{b}\xspace}

 \def\Pi      {\ensuremath{i}\xspace}

}

\def\bquark    {\ensuremath{\Pb}\xspace}

\def\kaon  {\ensuremath{\PK}\xspace}
  \def\Kbar  {\kern 0.2em\overline{\kern -0.2em \PK}{}\xspace}

\def\Kz    {\ensuremath{\kaon^0}\xspace}
\def\Kzb   {\ensuremath{\Kbar^0}\xspace}
\def\KzKzb {\ensuremath{\Kz \kern -0.16em \Kzb}\xspace}
\def\Kp    {\ensuremath{\kaon^+}\xspace}
\def\Km    {\ensuremath{\kaon^-}\xspace}

\def\KpKm  {\ensuremath{\Kp \kern -0.16em \Km}\xspace}

  \def\Dbar    {\kern 0.2em\overline{\kern -0.2em \PD}{}\xspace}
\def\D       {\ensuremath{\PD}\xspace}

\def\Dz      {\ensuremath{\D^0}\xspace}
\def\Dzb     {\ensuremath{\Dbar^0}\xspace}
\def\DzDzb   {\ensuremath{\Dz {\kern -0.16em \Dzb}}\xspace}
\def\Dp      {\ensuremath{\D^+}\xspace}
\def\Dm      {\ensuremath{\D^-}\xspace}

\def\DpDm    {\ensuremath{\Dp {\kern -0.16em \Dm}}\xspace}

  \def\Bbar    {\kern 0.18em\overline{\kern -0.18em \PB}{}\xspace}

  \def\Y#1S{\ensuremath{\PUpsilon{(#1S)}}\xspace}

\def\L {\ensuremath{\PLambda}\xspace}
\def\Lbar {\ensuremath{\kern 0.1em\overline{\kern -0.1em\PLambda}}\xspace}

\def\Lb      {\ensuremath{\L^0_\bquark}\xspace}

\def\to                 {\ensuremath{\rightarrow}\xspace}

\def\AT#1     {\ensuremath{A_{\mathrm{T}}^{#1}}\xspace}           

\def\C#1      {\ensuremath{\mathcal{C}_{#1}}\xspace}                       
\def\Cp#1     {\ensuremath{\mathcal{C}_{#1}^{'}}\xspace}                    
\def\Ceff#1   {\ensuremath{\mathcal{C}_{#1}^{\mathrm{(eff)}}}\xspace}        
\def\Cpeff#1  {\ensuremath{\mathcal{C}_{#1}^{'\mathrm{(eff)}}}\xspace}       
\def\Ope#1    {\ensuremath{\mathcal{O}_{#1}}\xspace}                       
\def\Opep#1   {\ensuremath{\mathcal{O}_{#1}^{'}}\xspace}                    

\newcommand{\tev}{\ensuremath{\mathrm{\,Te\kern -0.1em V}}\xspace}
\newcommand{\gev}{\ensuremath{\mathrm{\,Ge\kern -0.1em V}}\xspace}
\newcommand{\mev}{\ensuremath{\mathrm{\,Me\kern -0.1em V}}\xspace}
\newcommand{\kev}{\ensuremath{\mathrm{\,ke\kern -0.1em V}}\xspace}
\newcommand{\ev}{\ensuremath{\mathrm{\,e\kern -0.1em V}}\xspace}
\newcommand{\gevc}{\ensuremath{{\mathrm{\,Ge\kern -0.1em V\!/}c}}\xspace}
\newcommand{\mevc}{\ensuremath{{\mathrm{\,Me\kern -0.1em V\!/}c}}\xspace}
\newcommand{\gevcc}{\ensuremath{{\mathrm{\,Ge\kern -0.1em V\!/}c^2}}\xspace}
\newcommand{\gevgevcccc}{\ensuremath{{\mathrm{\,Ge\kern -0.1em V^2\!/}c^4}}\xspace}
\newcommand{\mevcc}{\ensuremath{{\mathrm{\,Me\kern -0.1em V\!/}c^2}}\xspace}

\def\gsim{{~\raise.15em\hbox{$>$}\kern-.85em
          \lower.35em\hbox{$\sim$}~}\xspace}
\def\lsim{{~\raise.15em\hbox{$<$}\kern-.85em
          \lower.35em\hbox{$\sim$}~}\xspace}

\def\tell1  {TELL1\xspace}
\def\ukl1   {UKL1\xspace}

\newcommand{\lbdnppi}{\ensuremath{\Lb\to D^0 p \pi^-}\xspace}

\title{Kernel density estimation of a multidimensional efficiency profile}

\author{Anton Poluektov$^{a,b}$\\
\llap{$^a$}University of Warwick,\\ 
Gibbet Hill Road, Coventry, CV4 7AL, UK\\
\llap{$^a$}Budker Institute of Nuclear Physics,\\ 
630090, Lavrentieva 11, Novosibirsk, Russia\\
E-mail: \email{Anton.Poluektov@cern.ch}
}

\abstract{
  Kernel density estimation is a convenient way to estimate the probability density of a distribution 
  given the sample of data points. However, it has certain drawbacks: 
  proper description of the density using narrow kernels needs large data samples, whereas 
  if the kernel width is large, boundaries and narrow structures tend to be smeared. 
  Here, an approach to correct for such effects, is proposed that uses an approximate 
  density to describe narrow structures and boundaries. The approach is shown to be well suited for 
  the description of the efficiency shape over a multidimensional phase space 
  in a typical particle physics analysis. An example is given for the five-dimensional 
  phase space of the \lbdnppi decay. 
}

\begin{document}

\section{Introduction}

A common task in particle physics analyses (as well as in many other areas)
is to obtain a functional representation of the probability density function (PDF)
$p(x)$ for a vector of variables $x$ from a data sample represented by a 
random set of values $x_i$. If the model behind the random process is known, 
a parametrised shape with a limited set of parameters is a good choice. 
This is, however, not always the case. Often the model is not known or is too complex
(or simply is not a subject of the study), in which case finding a 
good parametrisation can be difficult. Using a simplified parametrisation
leads to systematic biases, which have to be evaluated using alternative 
models. In the case of a multidimensional distribution, controlling the quality of 
a parametric fit can be difficult due to non-trivial correlations between 
the variables. 

On the other hand, non-parametric approaches, such as spline interpolation or
kernel density estimation do not require knowledge of the physics model 
behind the process and ensure the good quality of the PDF description by construction. 
These approaches, however, have their own shortcomings. 

This paper discusses the application of the kernel density estimation (KDE) technique 
for the description of a multidimensional PDF. Typical problems related to the 
use of traditional KDE are discussed: correction of boundary effects and 
description of narrow structures. A modification of the KDE procedure 
is proposed to solve these problems, which is shown to work well in a typical 
use case of high-energy physics analysis. A real-life example of the 
description of the reconstruction efficiency in the five-dimensional phase 
space of the \lbdnppi decay is given. A software package which implements the 
proposed technique is described. 

\section{Kernel density estimation basics}

Let us assume a random process which is characterised by a vector of variables $x$. 
The vector $x$ is in general multidimensional, but we will give an example with a scalar 
variable first. 
The distribution of variable $x$ follows a true PDF $P_{\rm true}(x)$ that we need to 
estimate. Let $x_i$ $(i=1\ldots N)$ be the data set which is sampled from this PDF. 
The kernel density estimator of the true PDF~\cite{Rosenblatt, Parzen} is
\begin{equation}
  P_{\rm KDE}(x) = \frac{1}{N}\sum\limits_{i=1}^{N} K(x-x_i). 
  \label{eq:kde}
\end{equation}
Here $K(x)$ is a kernel which is normalised to unity ($\int K(x) dx=1$). 
Various alternative forms can be used for $K(x)$. 
The notable property of the KDE technique is that the quality of the PDF description 
depends rather weakly on the actual kernel used. 
In the following, we will use Epanechnikov~\cite{epanech} form: 
\begin{equation}
  K(x) = \left\{
           \begin{array}{ll} 
             \frac{3}{4\sigma}\left(1-\frac{x^2}{\sigma^2}\right) & \mbox{for } x\in (-\sigma, \sigma) \\ 
             0              & \mbox{otherwise. } 
           \end{array} 
         \right.
  \label{eq:kernel}
\end{equation}
This kernel is optimal in some sense (it optimises a certain figure of merit called the
{\it asymptotic mean integrated square error, AMISE}~\cite{mise}) and is easy to calculate. 

Note that the resulting $P_{\rm KDE}(x)$ in the limit $N\to\infty$ with the fixed kernel width $\sigma$
is a convolution of the true PDF $P_{\rm true}(x)$ with the kernel $K(x)$. 
Thus, if the true PDF has relatively narrow structures (with characteristic width smaller than 
or equal to the kernel width), they will be smeared. 
Certainly, with larger samples, the kernel can be made narrower, so that 
asymptotically the estimated PDF $P_{\rm KDE}(x)$ will match the true one. For the limited sample size, 
however, one has to find a balance between the smearing of the PDF with large kernel widths, 
and the fluctuations of the estimated PDF with narrower kernels. 

\begin{figure}
  \centering
  \includegraphics[width=0.47\textwidth]{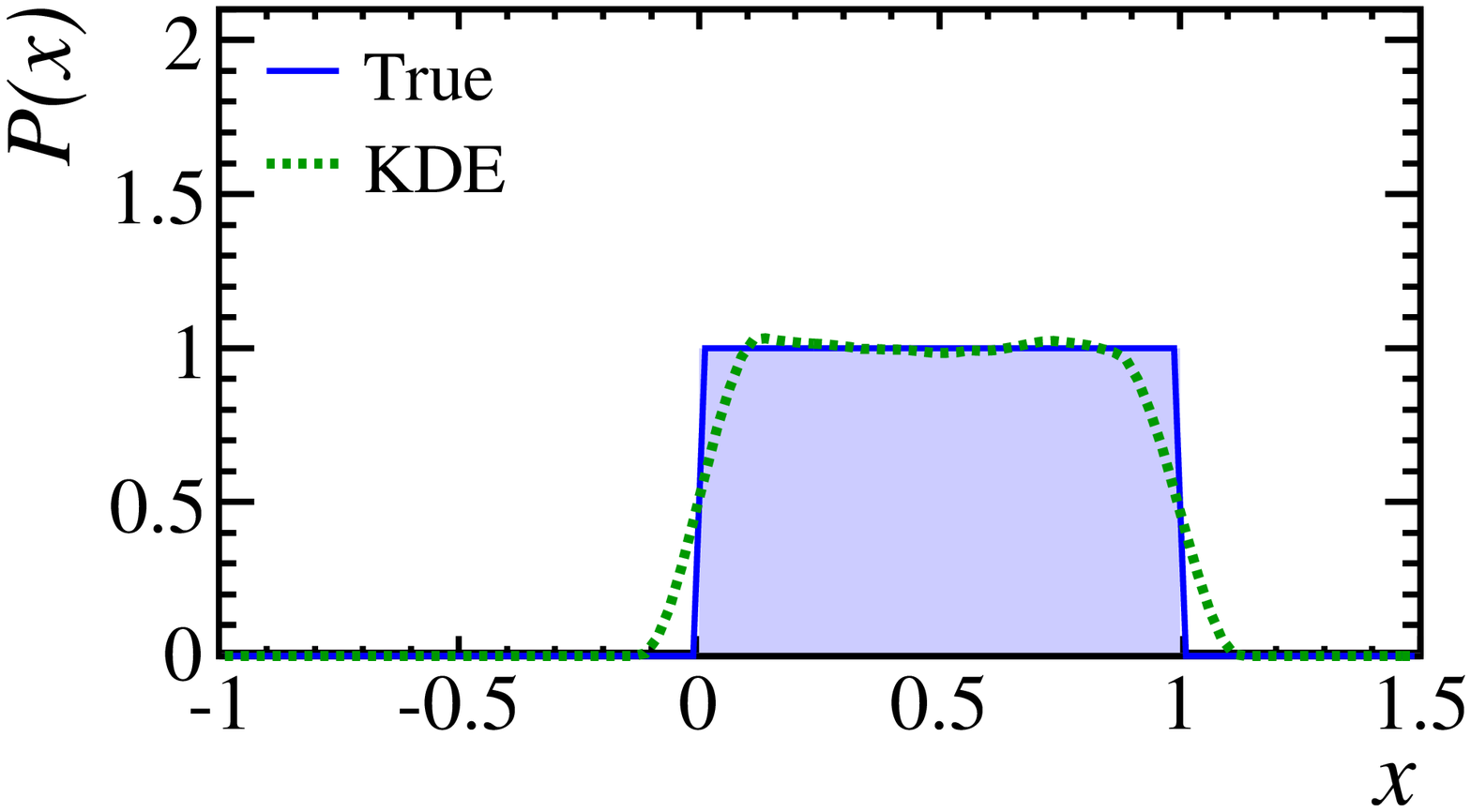}
  \put(-35,90){(a)}

  \includegraphics[width=0.47\textwidth]{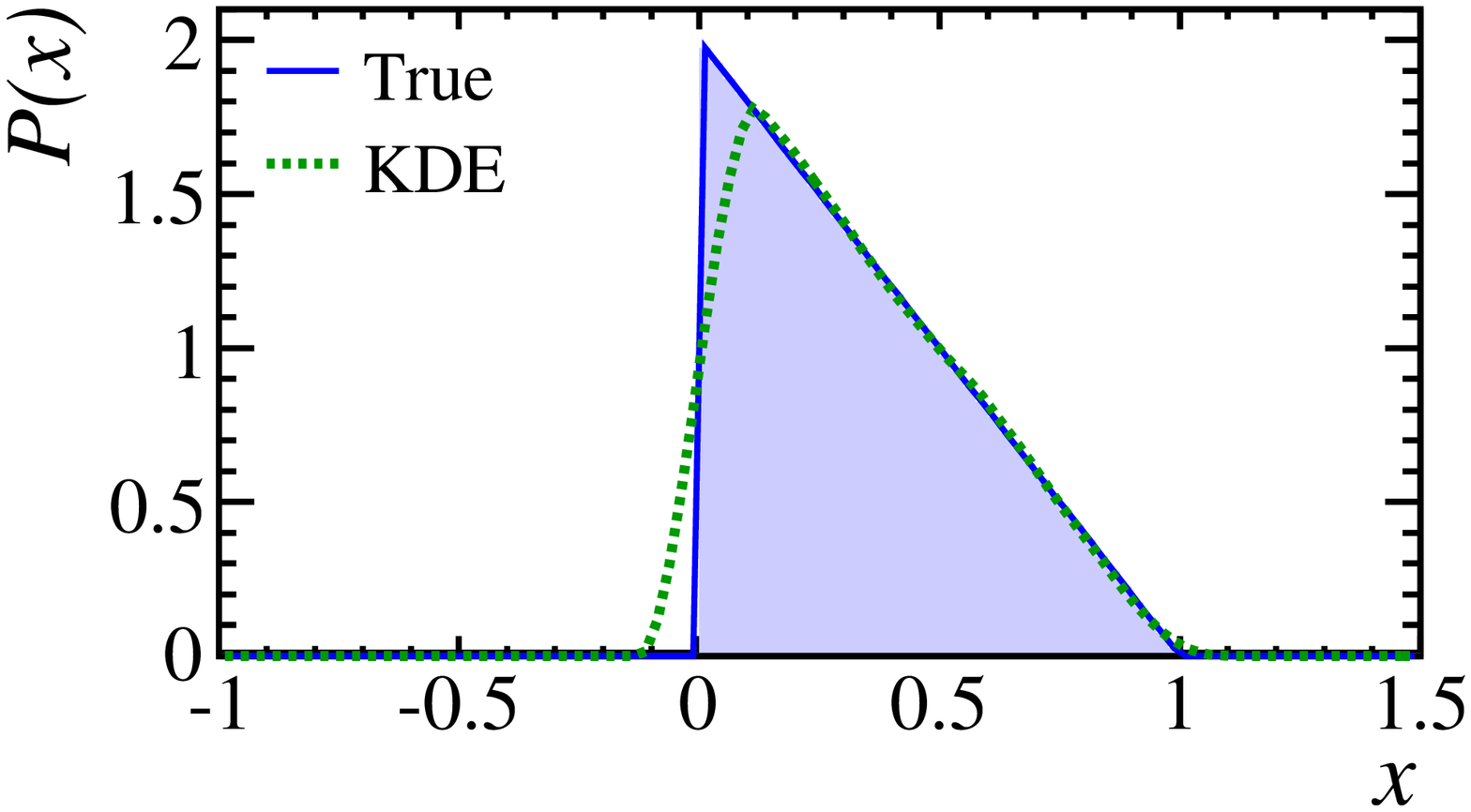}
  \put(-35,90){(b)}
  \hfill
  \includegraphics[width=0.47\textwidth]{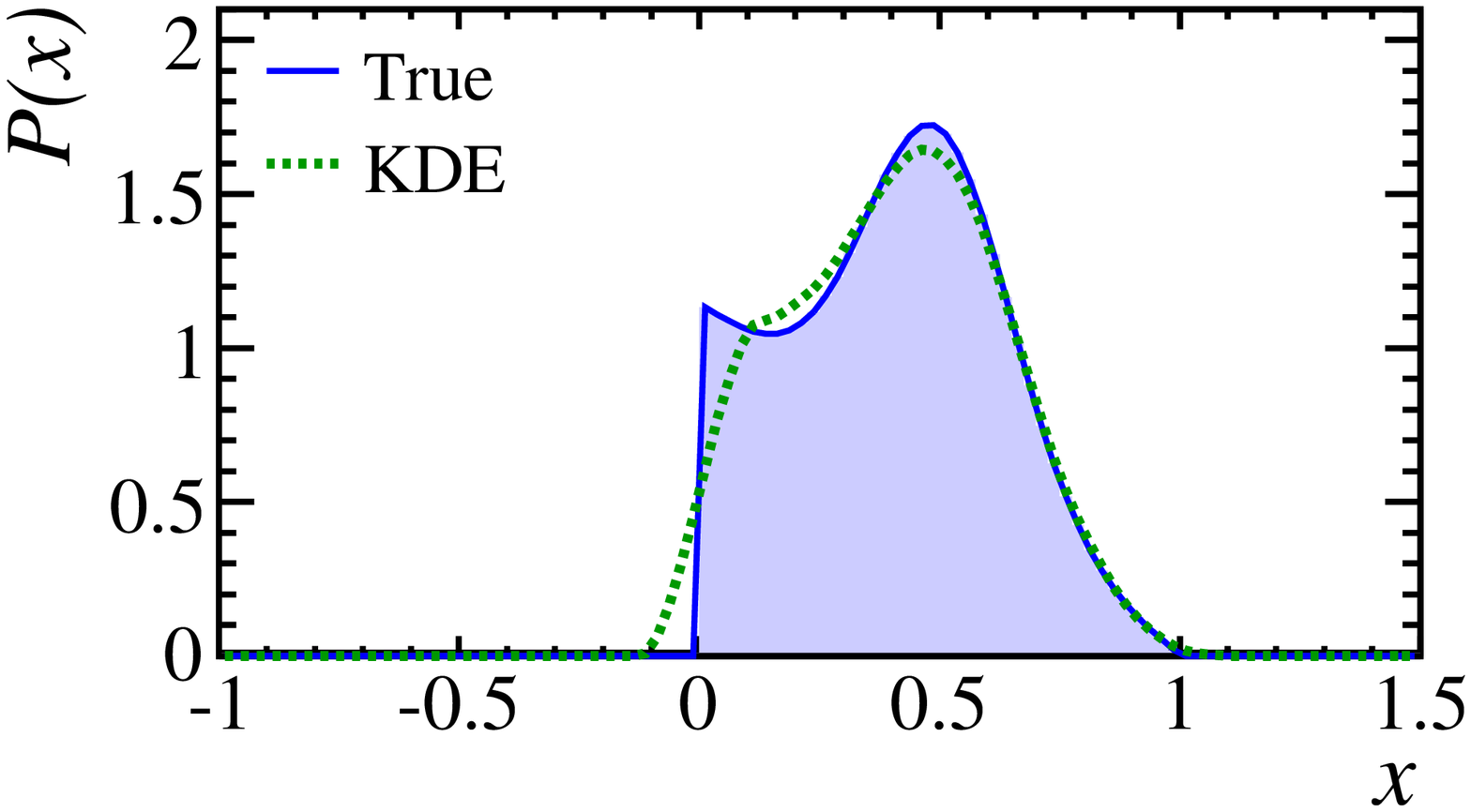}
  \put(-35,90){(c)}
  \caption{
    True PDF (filled area, blue solid line) and the result of 
    kernel density estimation (red dashed line), corresponding 
    to (a) uniform distribution, (b) linear distribution, and 
    (c) sum of a linear and a Gaussian distributions. 
  }
  \label{fig:uncorr}
\end{figure}

A typical problem of the KDE method is boundary effects. Direct use of Eq.~(\ref{eq:kde}) with a 
constant kernel results in an estimated PDF $P_{\rm KDE}(x)$ that falls off at the edges of the distribution
(Fig.~\ref{fig:uncorr}). If the function is defined for $x>0$ and is approximated by the linear function $1+\alpha x$ 
near the boundary $x=0$, the kernel PDF (\ref{eq:kde}) will give, up to the terms linear in $x$, 
\begin{equation}
  (1+\alpha x)\theta(x)\otimes K(x) = 
    \int\limits^{\sigma-x}_0 (1+\alpha y)\frac{3}{4\sigma}\left(1-\frac{x^2}{\sigma^2}\right) dy \simeq
    \left(\frac{1}{2} + \frac{3\alpha\sigma}{16}\right) + \left(\frac{\alpha}{2} + \frac{3}{4\sigma}\right) x, 
  \label{eq:kernel_bias}
\end{equation}
where $\theta(x)$ is the threshold function that represents the boundary at $x=0$: 
\begin{equation}
  \theta(x) = \left\{
           \begin{array}{ll}
             1 & \mbox{for } x\geq 0, \\
             0 & \mbox{for } x<0. 
           \end{array}
         \right.. 
\end{equation}
Thus, if the kernel width is small compared to the inverse derivative of the true PDF, $1/\alpha$, KDE 
gives a value which is twice lower at the boundary than the true PDF. The boundary effect 
is demonstrated for the one-dimensional case in Fig.~\ref{fig:uncorr} for a
uniform PDF, linear PDF, and the sum of a linear and a Gaussian PDF defined in the range $x\in (0, 1)$. 

Numerous methods have been proposed to correct for boundary effects: data reflection~\cite{Schuster}, 
pseudodata~\cite{Cowling, Hall}, kernel modification near the boundary~\cite{Jones, JonesFoster}, 
transformation of variables~\cite{Wand}. In some particular cases, such as in the case of a periodic function, 
the boundary correction can be performed in an exact way; since we consider a general case here, 
these techniques are not discussed. 
Normally the methods listed above work well with simple boundaries, 
such as in the one-dimensional case, or in a multidimensional case with rectangular 
boundaries. It is not straightforward to apply these methods in the cases of 
complex boundaries, such as for conventional Dalitz plots~\cite{dalitz}. 

\section{Boundary correction with approximation PDF}

A simple procedure is proposed here to correct for boundary effects, 
which works best if the approximate behaviour of the estimated function near the boundary is known.  
The procedure is illustrated below in a one-dimensional case, however it can also 
be applied in multiple dimensions with complex boundaries, as we will see later. 

Let us first consider the function $P_{\rm corr}(x)$ which is the ratio of the 
kernel PDF and the uniform function in the allowed region of $x$ values 
convolved with the 
kernel\footnote{Here and below the normalisation of the PDF is arbitrary. }: 
\begin{equation}
  P_{\rm corr}(x) = \left\{
           \begin{array}{ll}
             \frac{\sum\limits_{i=1}^{N} K(x-x_i)}{(U\otimes K) (x)} & \mbox{for } x\in X, \\
             0 & \mbox{otherwise}. 
           \end{array}
         \right.
  \label{eq:simple_corr}
\end{equation}
The function $U(x)$ is the constant function in the allowed range of $x$ and equals zero outside it: 
\begin{equation}
  U(x) = \left\{
           \begin{array}{ll}
             1 & \mbox{for } x\in X, \\
             0 & \mbox{otherwise}. 
           \end{array}
         \right.
\end{equation}

We can show how the PDF estimate of this form corrects for the boundary effect in the case of 
the linear PDF defined for $x>0$ near the boundary. 
Similarly to Eq.~(\ref{eq:kernel_bias}), in the limit $N\to \infty$ we obtain
\begin{equation}
  P_{\rm corr}(x)=\frac{(1+\alpha x)\theta(x)\otimes K(x)}{\theta(x)\otimes K(x)} = 
  \frac{\int\limits^{\sigma-x}_0 (1+\alpha y)\frac{3}{4\sigma}\left(1-\frac{x^2}{\sigma^2}\right)  dy}
       {\int\limits^{\sigma-x}_0 \frac{3}{4\sigma}\left(1-\frac{x^2}{\sigma^2}\right)  dy} \simeq
  \left(1+\frac{3\alpha\sigma}{8}\right) + \frac{7\alpha x}{16}
\end{equation}
up to the terms linear in $x$ and $\sigma$. 
Thus, the approach given by Eq.~(\ref{eq:simple_corr}), unlike the uncorrected KDE, 
is asymptotically unbiased at the boundary for small kernel widths ($\alpha\sigma \ll 1$). 
However, if the kernel width is large ($\alpha\sigma\sim 1$), as it is likely to
be in the case of a multidimensional PDF, bias does occur. 

\begin{figure}
  \includegraphics[width=0.47\textwidth]{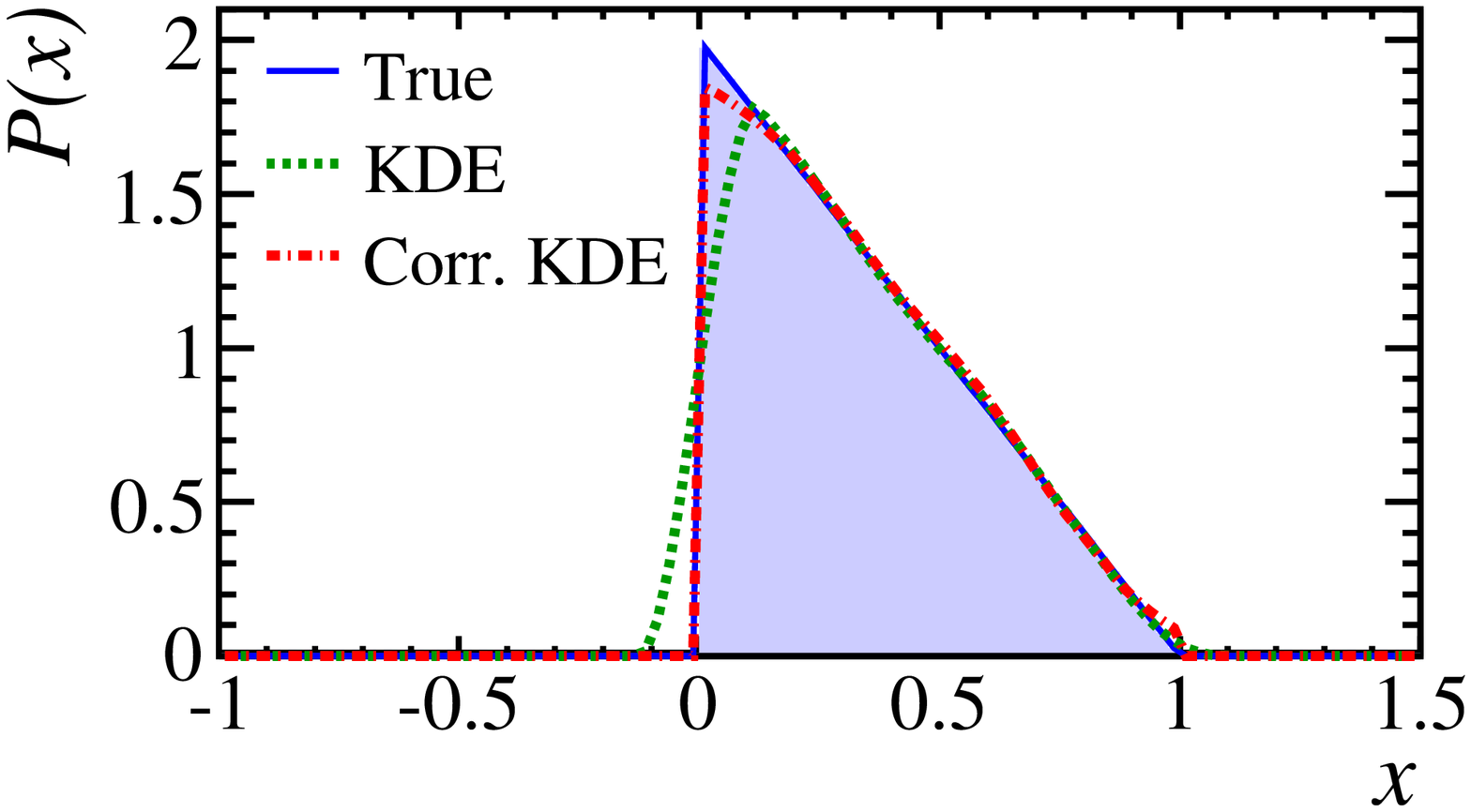}
  \put(-35,90){(a)}
  \hfill
  \includegraphics[width=0.47\textwidth]{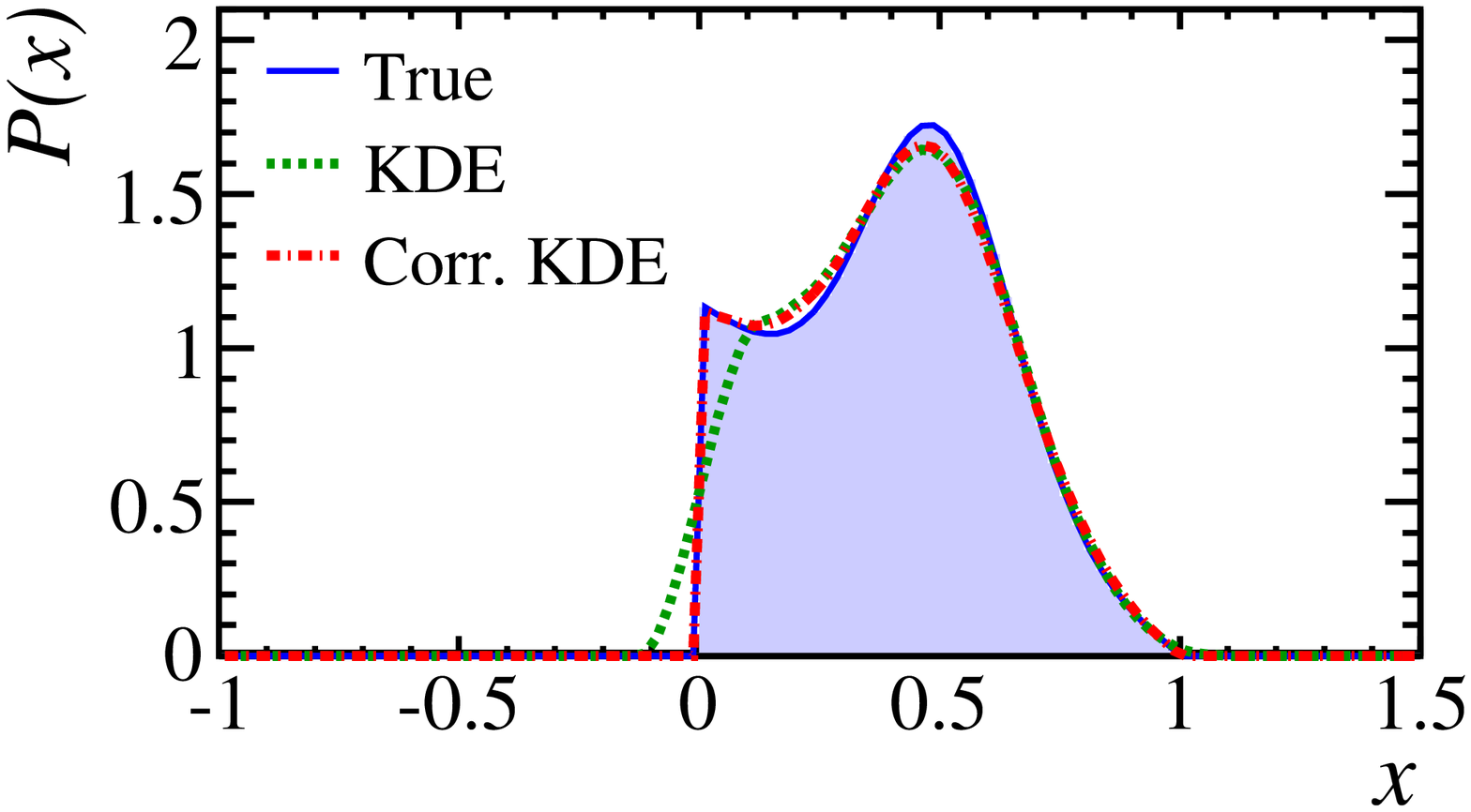}
  \put(-35,90){(b)}
  \caption{(a) True PDF (filled area, blue solid line), the result of 
    kernel density estimation (green dashed line), and of the 
    KDE corrected as in 
    Eq.~(3.1) (red dash-dotted line). 
    (b) The same for the sum of a linear and a Gaussian 
    distribution, with correction applied as in 
    Eq.~(3.4)
    using a linear PDF as an approximation.}
  \label{fig:corr}
\end{figure}

The effect here is similar to that observed when the data reflection technique 
is used for boundary correction. In both cases, the kernel estimate is unbiased for 
small kernel widths, but the derivative of the estimated PDF at the boundary is 
underestimated: it equals zero by construction for the data reflection technique, 
and equals $7\alpha/16$ in the case above (so the bias is strictly smaller than for the 
data reflection technique).  The effect of the boundary correction using Eq.~(\ref{eq:simple_corr}) 
is illustrated in Fig.~\ref{fig:corr}(a) for the linear true PDF. 

    

Note that if the true PDF is flat at the boundary ({\it i.e.} if $\alpha=0$), 
expression (\ref{eq:simple_corr}) does not lead to a bias even for 
the finite kernel width. One can now modify this expression in such a way that
the estimate be unbiased not only for $\alpha=0$, but in general for any {\it a-priori}
known behaviour of the true PDF near the boundary. Let us assume that the true PDF 
is described by the approximation function $F(x)$ in the vicinity of the boundary ($x<\sigma$). 
Consider the expression of the form
\begin{equation}
  P_{\rm approx}(x) = \frac{\sum\limits_{i=1}^{N} K(x-x_i)}{(F\otimes K) (x)}\times F(x). 
  \label{eq:approx_corr}
\end{equation}
In the limit of a large data sample ($N\to\infty$), for a fixed kernel width $\sigma$ this expression 
is equal to the approximation PDF $F(x)$ near the boundaries, since 
$\sum\limits_{i=1}^{N} K(x-x_i)\simeq (F\otimes K)(x)$. 
In the inner regions, where in the absence of narrow structures $(F\otimes K)(x) \simeq F(x)$, 
one obtains $P_{\rm approx}\simeq P_{\rm KDE}$. 
In the expression above we assume that the allowed region $X$ of the values of variable $x$ is defined 
by the function $F(x)$ such that $F(x)=0$ for $x\notin X$. 

As an example, we show the effect of PDF correction using Eq.~(\ref{eq:approx_corr}) in Fig.~\ref{fig:corr}(b). 
Here we estimate the PDF which is the sum of a linear and a Gaussian distributions, 
using the linear function as an approximation PDF. 

The use of Eq.~(\ref{eq:approx_corr}) is not only limited to the boundary correction. 
Other narrow structures, which otherwise would be smoothed out by the kernel,
can also be described by the approximation PDF $F(x)$. In general, if the PDF we want to 
parametrise can be factorised as $P_{\rm true}(x) = f(x)F(x)$, where $F(x)$ is a known 
function which possibly contains narrow structures, while $f(x)$ is slowly varying
compared to the kernel width, it can be efficiently parametrised by this technique. 
In other words, the kernel density estimator in Eq.~(\ref{eq:approx_corr}) 
describes the {\it relative} variations of the PDF with respect to the approximation function $F(x)$. 
We will therefore refer to this approach as the {\it relative kernel density estimation} technique. 

The proposed approach where a nonparametric correction is applied on top of an approximation PDF 
can be effectively applied in many cases for multidimensional PDFs. In general, 
obtaining PDF estimation for a multidimensional function requires large data samples. 
Often, however, the PDF can be approximately factorised as a product of several PDFs with reduced 
number of dimensions. In the approximation approach, these PDFs can be described again using 
the same kernel density technique with a kernel of narrower width. The approximation PDF can 
then be taken as the product of these PDFs, while the residual correction can be applied using 
the proposed approach with a larger kernel width. 

A typical particle physics analysis uses a description of the efficiency profile 
from full detector simulation which is an extremely CPU-consuming task. Using the 
proposed approach can help reduce the required simulation sample. For instance, 
an approximation efficiency function can be obtained from a large sample of data obtained 
using simplified, but fast simulation, while the final correction can come from the 
sample of full detector simulation. 

\section{Practical considerations}

Direct usage of the relative KDE approach as in Eq.~(\ref{eq:approx_corr}) in the fit of 
experimental data can in practice be slow. For each data point $x$ one has to calculate the 
convolution of the approximation function $F(x)$ with the kernel. In general, if the 
convolution is not expressed analytically, one has to use Monte-Carlo (MC) integration or a similar 
numerical technique to calculate it. It could be more effective to store the calculated values of 
the estimated PDF on a multidimensional grid, and use linear interpolation between the grid nodes 
to calculate the values of the function in the data fit. 

Calculation of the function using linear interpolation between the nodes of a rectangular 
grid in the presence of a (in general, curved) physical boundary, however, is not straightforward. 
To obtain the value of the function for points close to the boundary, 
the function has to be calculated for nodes which are outside of the boundary. 
Fortunately, the relative KDE approach allows extrapolation of the function $P(x)$ 
beyond the boundary under certain conditions: 
the density expressed by Eq.~(\ref{eq:approx_corr}) can be calculated for the points 
beyond the boundary if the distance between the extrapolated point and the boundary 
is small compared to the kernel width. 

Interpolation can be applied directly to the result of relative KDE estimation (\ref{eq:approx_corr}), 
{\it i.e.}
\begin{equation}
  P_{\rm interp}(x) = Bin(P_{\rm approx}(x)) = Bin\left(\frac{\sum\limits_{i=1}^{N} K(x-x_i)}{(F\otimes K) (x)}\times F(x)\right), 
  \label{eq:binned}
\end{equation}
where the functor $Bin(f)$ denotes multilinear interpolation of the function $f$ over the rectangular grid. 
However, calculating $P_{\rm interp}(x)$  in this way is computationally intensive. For $k$ grid points, 
a vector of data $x_i$ of size $n$ and $N$ points in the toy MC convolution vector, the calculation 
takes time proportional to $k(n+N)$, and needs both the data vector and convolution vector
to be kept in memory. An alternative approach is to perform linear interpolation for both the numerator 
and denominator of Eq.~(\ref{eq:approx_corr}): 
\begin{equation}
  P_{\rm interp}(x) = \frac{Bin\left(\sum\limits_{i=1}^{N} K(x-x_i)\right)}{Bin\left((F\otimes K) (x)\right)}\times F(x). 
  \label{eq:binned2}
\end{equation}
In that case one has to loop though the data and convolution vectors only once, and thus they don't 
need to be kept in memory. In addition, for each point of the data vector, one has to loop through 
not all the nodes of the grid, but only over those nodes which are less than a kernel width apart from 
the point $x_i$. 

None of the expressions above assume that the kernel is constant over the phase space. 
It is therefore possible to apply an adaptive kernel technique, where the kernel width is a function 
of the PDF value. In practice, this would result in an iterative procedure, where the kernel width in 
each iteration depends on the PDF value calculated in the previous iteration. Such an approach could be 
useful for PDFs containing sharp peaks. 

The conventional kernel density estimator works naturally with weighted data sets, where each event $x_i$
has its individual weight $w_i$. In that case, Eq.~(\ref{eq:kde}) becomes
\begin{equation}
  P_{\rm KDE}(x) = \frac{\sum\limits_{i=1}^{N} w_i K(x-x_i)}{\sum\limits_{i=1}^{N} w_i}. 
  \label{eq:weighted_kde}
\end{equation}
The relative KDE technique is also easily generalised to deal with  
weighted data sets in a similar way, by using the weighted sum in the numerator of Eq.~(\ref{eq:approx_corr}). 

\section{{\tt Meerkat} package}

The technique described here has been implemented in the software package called 
{\tt Meerkat} (which stands for 
``Multidimensional Efficiency Estimation using the Relative Kernel Approximation Technique''), 
available at HepForge~\cite{meerkat}. 
The package provides a set of {\tt C++} classes for the estimation of a PDF of any (reasonable) 
dimensionality defined over a phase space which is the combination of Dalitz plot phase spaces, 
one-dimensional phase spaces, and phase spaces with limits defined in a functional form. 
The code depends on the ROOT~\cite{root} framework. 

The classes of the {\tt Meerkat} library are derived from one of two abstract classes, \\
{\tt AbsPhaseSpace} which describes the phase space, and {\tt AbsDensity} which defines the 
interface for the PDF classes. The classes derived from {\tt AbsPhaseSpace} are as follows: 
\begin{itemize}
  \item {\tt OneDimPhaseSpace} describes a one-dimensional 
   phase space, which is just the finite range of values of the single scalar variable. 

  \item {\tt DalitzPhaseSpace} is the two-dimensional phase space of the Dalitz plot, 
   the kinematic phase space of the decay of a particle into three other particles. 
 
  \item {\tt ParametricPhaseSpace} allows to define an arbitrary phase space 
   where the boundaries (lower and upper limit) of one variable are functions 
   of variables over another phase space. 

  \item {\tt CombinedPhaseSpace} describes a combination (direct product) 
   of several component phase spaces. 
\end{itemize}
The classes derived from {\tt AbsDensity} are: 
\begin{itemize}
  \item {\tt UniformDensity} is a uniform density over arbitrary phase space. 
  
  \item {\tt FormulaDensity} represents a fixed density which is defined in 
        a functional form. Both {\tt UniformDensity} and {\tt FormulaDensity}
        can be used as the approximation PDFs for the kernel density estimator. 

  \item {\tt KernelDensity} implements the unbinned relative KDE procedure defined by 
        Eq.~(\ref{eq:approx_corr}). 
 
  \item {\tt BinnedDensity} defines a PDF which is the result of multilinear 
        interpolation of a PDF over arbitrary phase space ({\it i.e.} it implements the $Bin(f)$ 
        functor as in Eq.~(\ref{eq:binned})). 

  \item {\tt BinnedKernelDensity} implements the binned approach as in Eq.~(\ref{eq:binned2}). 
        It is much more efficient from the point of view of computational time and memory usage 
        than\\ 
        {\tt KernelDensity}, 
        and is thus recommended, except for the special cases when the binning introduces 
        unacceptable distortions of the approximation function. 

  \item {\tt AdaptiveKernelDensity} implements the same approach as\\ {\tt BinnedKernelDensity}, 
        but with a variable kernel width which is a function of the value of another PDF
        ({\it e.g.} of the PDF estimated at the previous iteration). 
        Both\\ {\tt BinnedKernelDensity} and {\tt AdaptiveKernelDensity} classes can work with \\
        weighted data sets. 
        
  \item {\tt FactorisedDensity} is the density which is simply a product of two or more 
        densities of different (sets of) variables. {\tt FactorisedDensity} can, for instance, 
        be used as approximation PDF for {\tt BinnedKernelDensity}. 
\end{itemize}
The code also contains several examples of usage of these classes. 

\section{Application to \lbdnppi phase space}

One possible use case of the procedure proposed here is the description of efficiency, 
background, and similar quantities over multidimensional kinematic phase space in 
high-energy physics processes.\footnote{In the following, we will refer to all these 
quantities as ``probability densities'' while, strictly speaking, these functions are 
only proportional to the PDFs estimated from the samples of random points. In our task 
of estimating the multidimensional shape of these functions, correcting 
the common normalisation is straightforward and is thus not discussed. } 
In a multidimensional case, the size of the data sample to 
be used to parametrise these quantities becomes a critical issue. It is often assumed that 
some or all the kinematic variables describing the decay are uncorrelated and thus the PDF 
can be represented in a factorised form. This effectively reduces the dimensionality and the size of the 
data sample needed. However, possible deviation from the factorisation assumption 
can bias the measurement and has to be treated as a systematic uncertainty. In cases when 
the correlations between the variables are small, or at least not changing rapidly, the 
approach proposed here can be efficiently used. 

Below we illustrate this approach in the real-life application to the description of the 
selection efficiency of the baryonic decay \lbdnppi which is being studied
at the LHCb experiment~\cite{Aaij:2013pka}. This is a three-body 
decay where both the initial and the final states contain spin-$1/2$ particles ($\Lb$ and 
the proton). In the general case, the kinematic distribution is fully described by five variables. 
One can choose {\it e.~g.}, two squared invariant masses of the pairs of final state particles, $M^2(D^0p)$ and 
$M^2(p\pi)$, and three angles which describe the orientation of the decay plane in a certain reference frame, 
$\theta_p$, $\phi_p$ and $\phi_{D\pi}$. 

To validate the proposed procedure we need a large sample of decays, therefore we use a simplified 
simulation of the LHCb detector response. Our simulation reflects the forward geometry 
of the LHCb detector~\cite{lhcb}, which instruments the region close to the beam line of the LHC.
We use coordinates where the $z$ axis corresponds to the beam line, and the instrumented 
region is at $z>0$. Decays of $\Lb$ baryons are generated using the EvtGen~\cite{evtgen} package
with uniform distribution in the $D^0p\pi$ phase space with $D^0$ decaying to $K^-\pi^+$. 
$\Lb$ baryons are generated unpolarised, with uniform 
distribution in pseudorapidity $\eta = -\log\tan\theta/2$ in the range $1.5<\eta<5.5$ (where 
$\theta$ is the polar angle in the chosen frame), are distributed uniformly in azimuthal angle $\phi$
and have exponential distribution of the transverse (with respect to the beam line) momentum
\begin{equation}
  f(p_{\rm T}) \propto \exp(-p_{T\rm }/p^0_{\rm T}), 
\end{equation}
where $p^0_{\rm T}=5$ GeV$/c$. To simulate the tracking system and trigger response and the offline 
selection, the following requirements are applied to the decay products of the \Lb: 
\begin{itemize}
  \item All charged particles in the final state are required to have pseudorapidity in 
        the range $2<\eta<5$ (tracking system acceptance), transverse momentum $p_{\rm T}>0.15$ GeV$/c$ 
        (offline selection to reduce background) and momentum $p<100$ GeV$/c$ 
        (requirement of reliable particle identification). 
  \item Pions and kaons in the final state have to satisfy $p>5$ GeV$/c$, while protons 
        are required to have $p>10$ GeV$/c$ (particle identification thresholds). 
  \item At least one particle in
        the final state is required to have transverse momentum $p_{\rm T}>3.5$ GeV$/c$
        (hardware trigger selection). 
\end{itemize}

In this example we will illustrate the strong point of the proposed approach that is 
the possibility to gradually, in several steps, increase the dimensionality of the PDF estimator, 
by using the factorised approximation PDF from the previous step, and correcting for correlations 
using the relative KDE technique. 

We will see how well the 5D distribution of the \lbdnppi decay can be estimated using 
a sample of limited size of $10^5$ decays. We make use of a large sample of simulated data 
($10^8$ events) to obtain the reference (``true'') distribution of the efficiency with a relatively narrow 
kernel, and then use subsamples of $10^5$ events to evaluate the quality of the PDF estimation 
in comparison with the reference PDF. The kernel width for the KDE with a limited data sample has 
to be optimised to reach a balance between the statistical fluctuations of the estimated PDF 
(which increase for narrower widths) and systematic bias due to smearing of the structures 
(which becomes critical for wider kernels). The following figure of merit is used to evaluate 
the quality of the estimation: 
\begin{equation}
  Q = \sqrt{\sigma^2 + \Delta^2}, 
\end{equation}
where 
\begin{equation}
  \sigma = \langle RMS_i(F_i(x_{\alpha}) ) \rangle_{\alpha}
\end{equation}
is the variance $\sigma_{\alpha}$ of the PDF values in the node $\alpha$ of the rectangular grid, 
averaged over all nodes within the allowed phase space, and 
\begin{equation}
  \Delta = RMS_{\alpha}( \langle F_i(x_{\alpha}) \rangle_{i})
\end{equation}
is the RMS (calculated over the nodes $\alpha$ of the rectangular grid) 
of the average bias of estimated PDF values with respect to the reference value. 
Here $\langle\ldots\rangle_{i}$ and $\langle\ldots\rangle_{\alpha}$ denote averaging 
over the subsample and grid node, respectively, while $RMS_i(\ldots)$ and $RMS_{\alpha}(\ldots)$ 
mean taking the RMS value. 

\subsection{Estimation of the Dalitz plot density}

\label{sec:dalitz}

Before we start to deal with the full 5D phase space, we will consider the estimation of 
PDFs with reduced dimensionality. These PDFs will then be used as components of the 5D approximation 
PDF. We start with the estimation of the 2D distribution in Dalitz plot variables 
$M^2(D^0p)$ and $M^2(p\pi)$. The characteristic property of this distribution is that the allowed 
phase space in these variables has a non-trivial shape, with the range of one variable depending on 
the value of the other. 

\begin{figure}
  \includegraphics[width=0.47\textwidth]{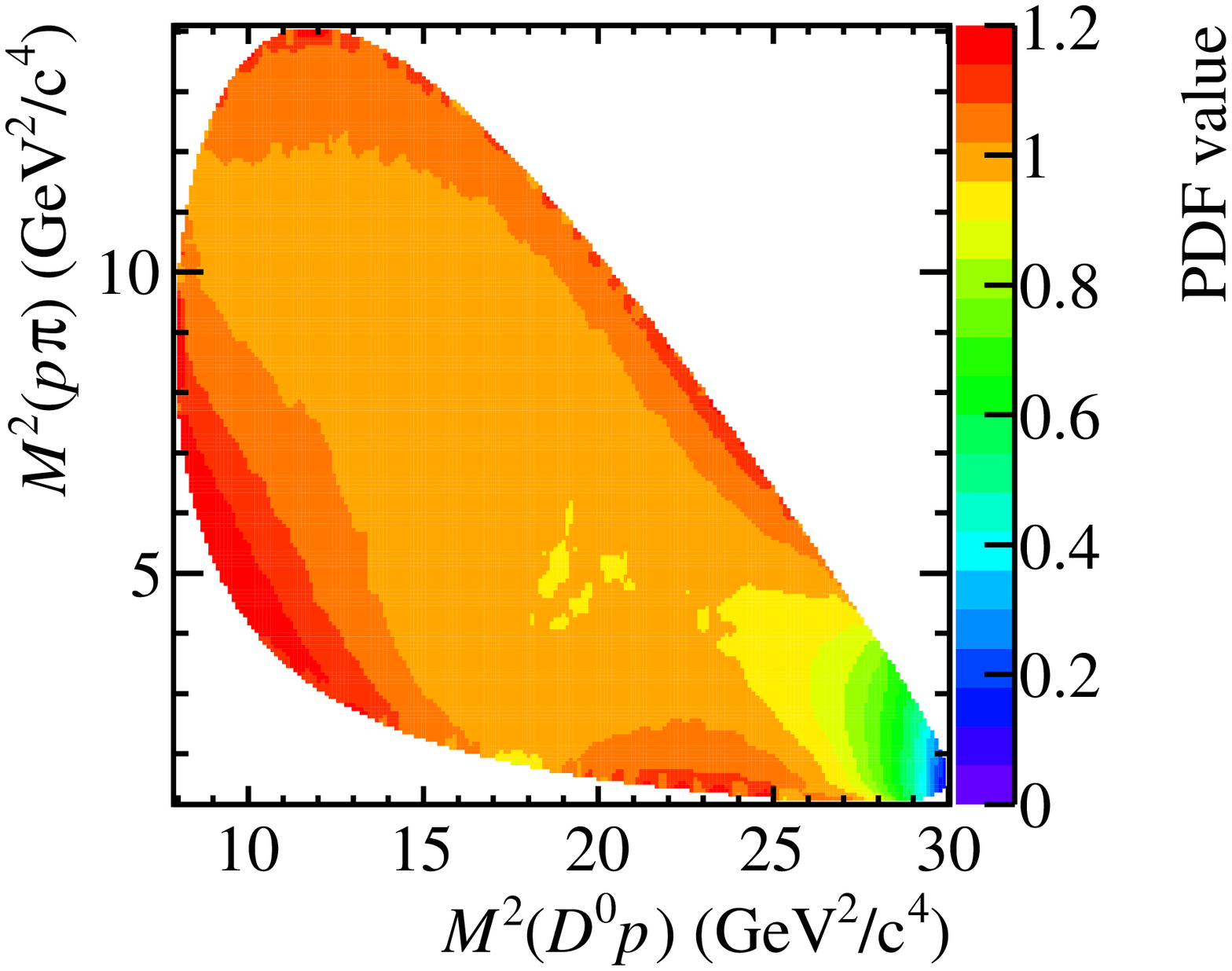}
  \put(-65, 140){(a)}
  \includegraphics[width=0.47\textwidth]{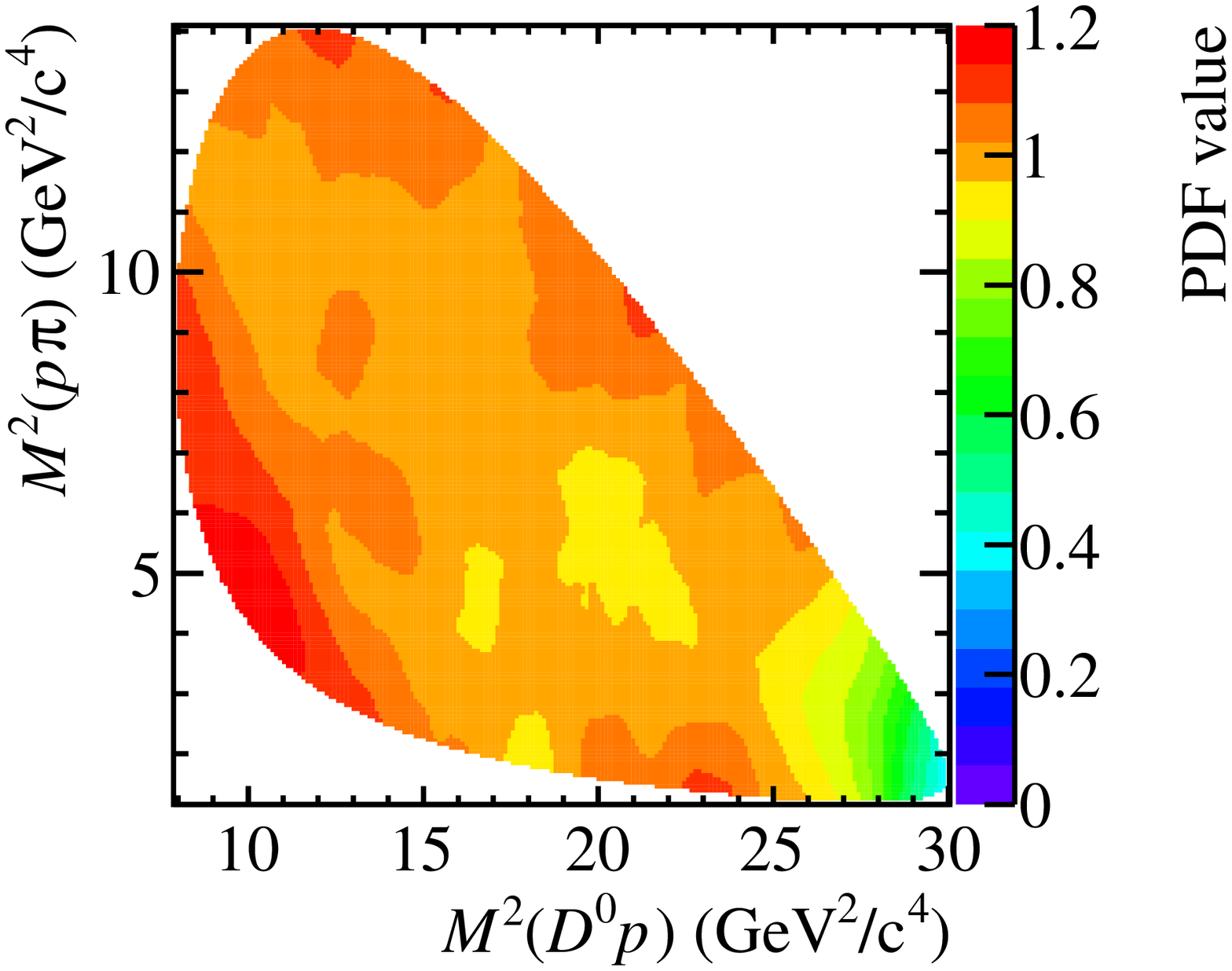}
  \put(-65, 140){(b)}
  
  \includegraphics[width=0.47\textwidth]{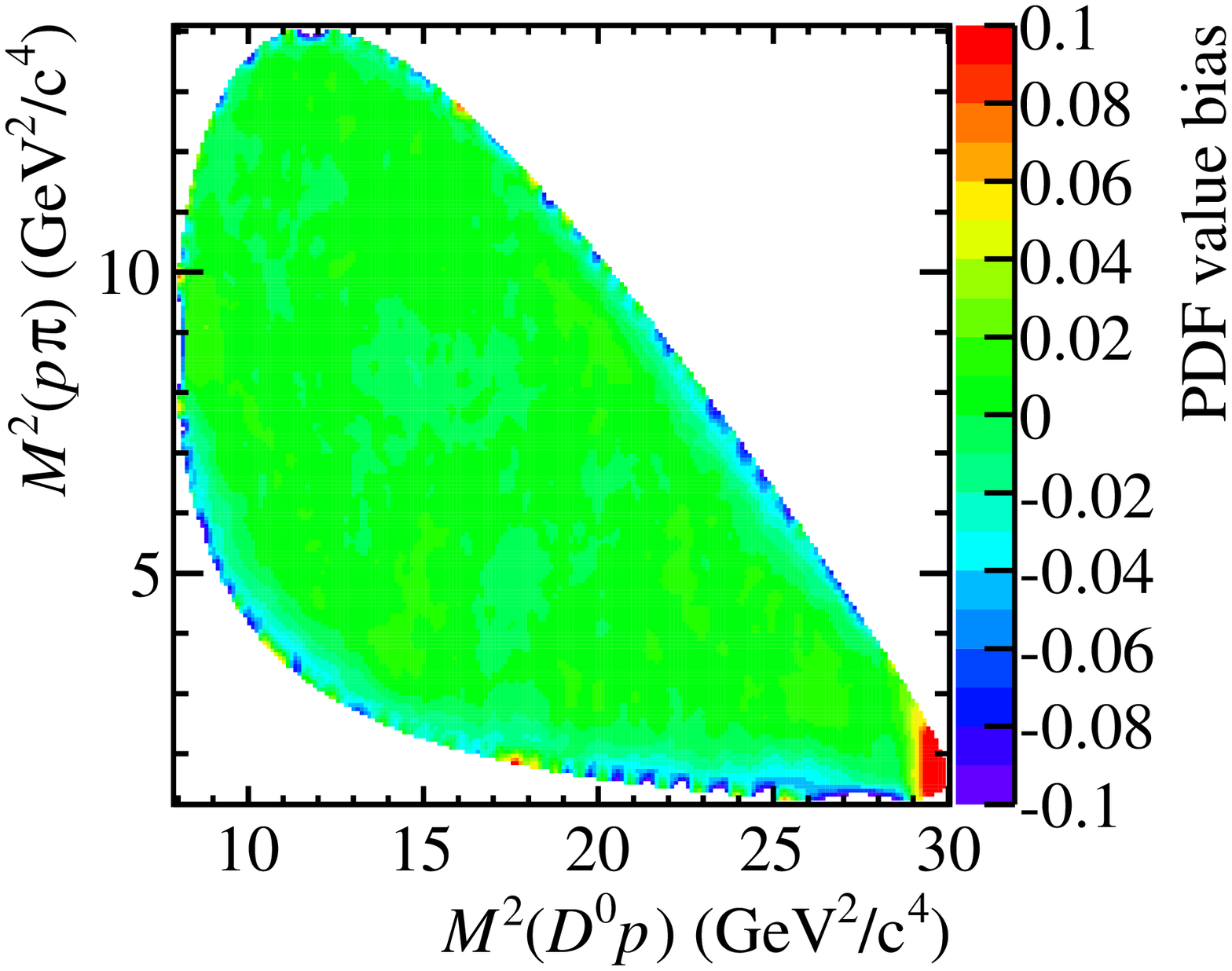}
  \put(-65, 140){(c)}
  \includegraphics[width=0.47\textwidth]{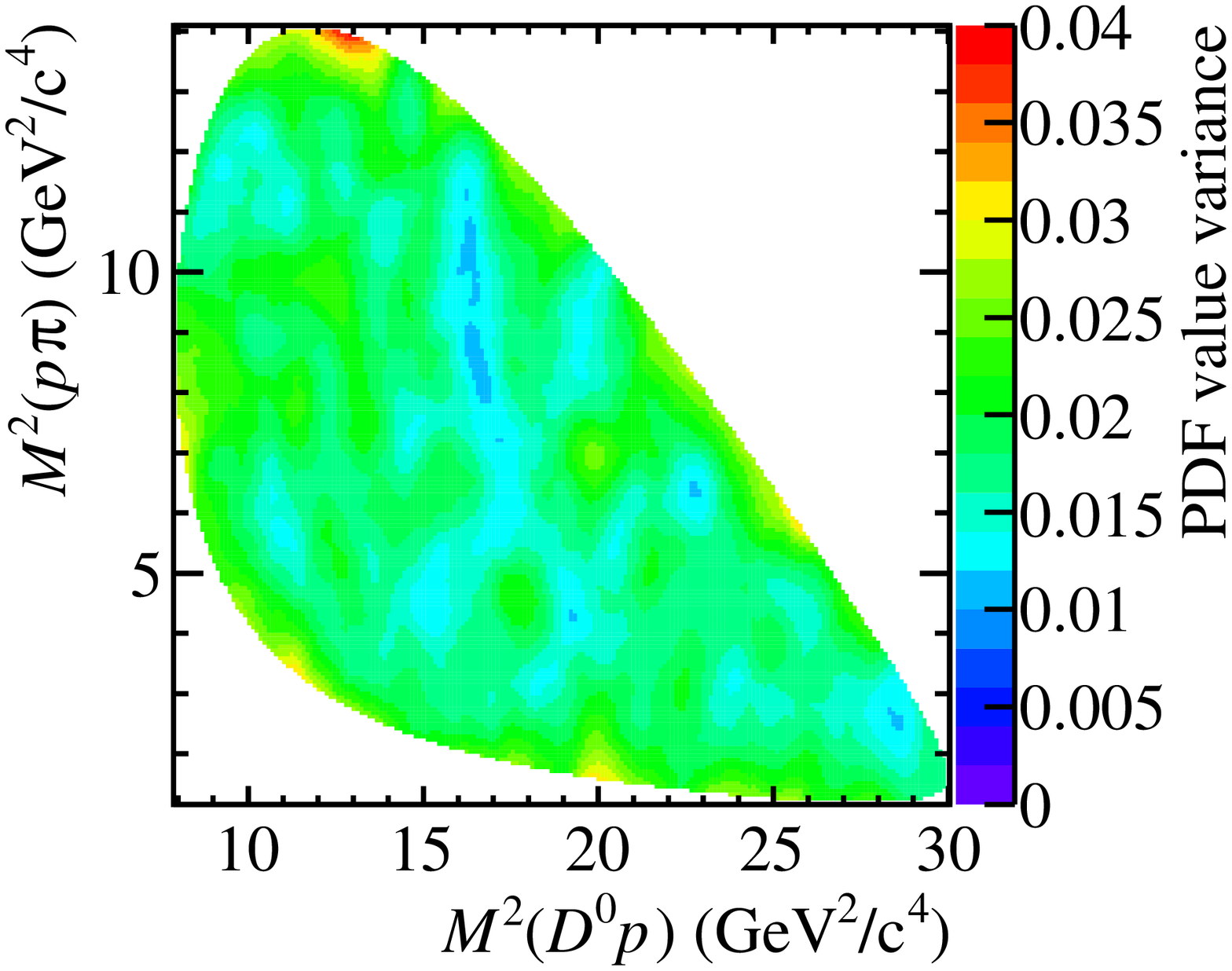}
  \put(-65, 140){(d)}
  \caption{Results of the estimation of the efficiency of $\lbdnppi$ decays across the Dalitz plot. 
           (a) Reference 
           density obtained with narrow kernel and (b) an example of density obtained with $10^5$
           sample size and kernel width of $2$ GeV$^2/c^4$. (c) Average bias and (d) 
           variance of estimated PDF values as a function of Dalitz plot position. }
  \label{fig:dalitz}
\end{figure}

Reference density over the Dalitz plot, obtained with a narrow kernel of 0.3 GeV$^2/c^4$ width 
in both $M^2(D^0p)$ and $M^2(p\pi)$ variables from the sample of $10^8$ simulated decays, 
is shown in Fig.~\ref{fig:dalitz}~(a). Figure~\ref{fig:dalitz}~(b) shows an example of the density 
obtained from a sample of $10^5$ events with a kernel width of 1.5 GeV$^2/c^4$. To allow direct 
comparison, all PDFs are normalised such that the average PDF value over the phase space equals 
to one. As a result of 
running the PDF estimation on 50 samples of $10^5$ events, we find the bias and variance of the PDF 
value as a function of Dalitz plot position. These are shown in Figs.~\ref{fig:dalitz} (c) and (d), 
respectively. The quantities $\Delta$ and $\sigma$ are calculated from these distributions. 
The values of RMS bias $\Delta$ and average variance $\sigma$, as well as the overall quality of 
the density estimation $Q$, are given in Table~\ref{tab:dalitz} for three different kernel widths. 
The width of 1.5 GeV$^2/c^4$ gives the best value of $Q=0.0165$. This can be interpreted as that the 
average precision of the PDF value estimation (including statistical and systematic effects) is 1.65\%. 

\begin{table}
  \caption{Results of the estimation of the Dalitz plot efficiency profile for the \lbdnppi decays. 
           Systematic ($\Delta$), variance ($\sigma$) terms and the overall figure 
           of merit $Q$ are given for different kernel widths. }
  \label{tab:dalitz}
  \begin{center}
  \begin{tabular}{l|ccc}
    Kernel width, MeV$^2/c^4$ & Bias $\Delta$ & Variance $\sigma$    & $Q$ \\
    \hline
    1.0                    & 0.0084   & 0.0159      & 0.0179 \\
    1.5                    & 0.0125   & 0.0107      & 0.0165 \\
    2.0                    & 0.0173   & 0.0081      & 0.0191 \\ 
  \end{tabular}
  \end{center}
\end{table}

\subsection{Estimation of the angular density}

\label{sec:angular}

The second component of the \lbdnppi phase space is the angular part, which is described by three angles. 
The variables used to describe it are defined in the rest frame of the decaying \Lb, with the $x'$ axis 
given by its direction in the laboratory frame, the polarisation axis $z'$ given by the cross product of the 
beam and $x'$ axes, and the $y'$ axis by the cross product of the $z'$ and $x'$ axes. The three variables are
cosine $\cos\theta_p$ of the polar angle and the azimuthal angle $\phi_p$ of the proton momentum in this reference frame, 
and the angle $\phi_{Dp}$ between the $D^0\pi$-plane and the plane formed by the proton and polarisation axis. 

\begin{figure}
  \includegraphics[width=0.47\textwidth]{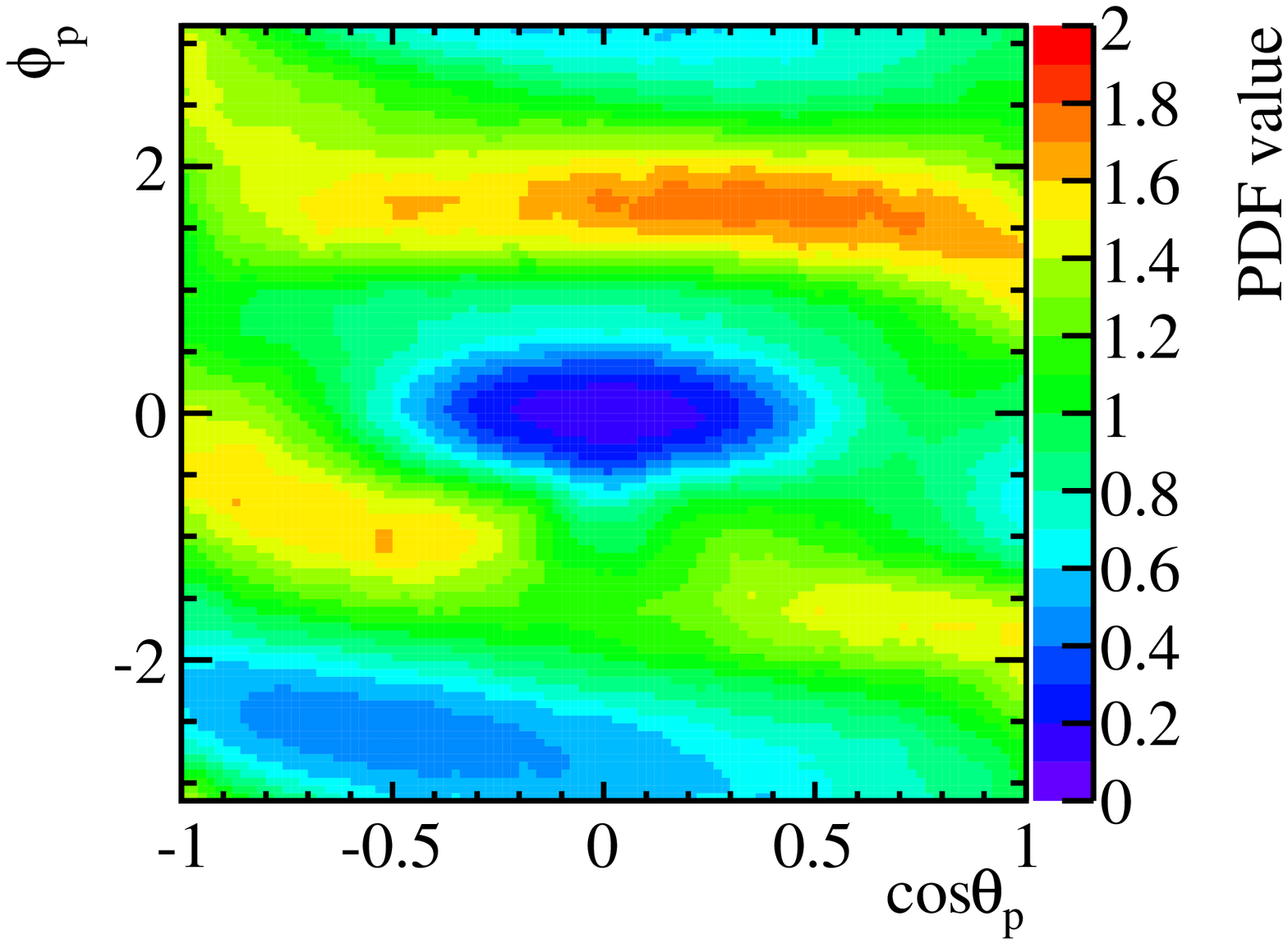}
  \put(-60, 130){(a)}
  \includegraphics[width=0.47\textwidth]{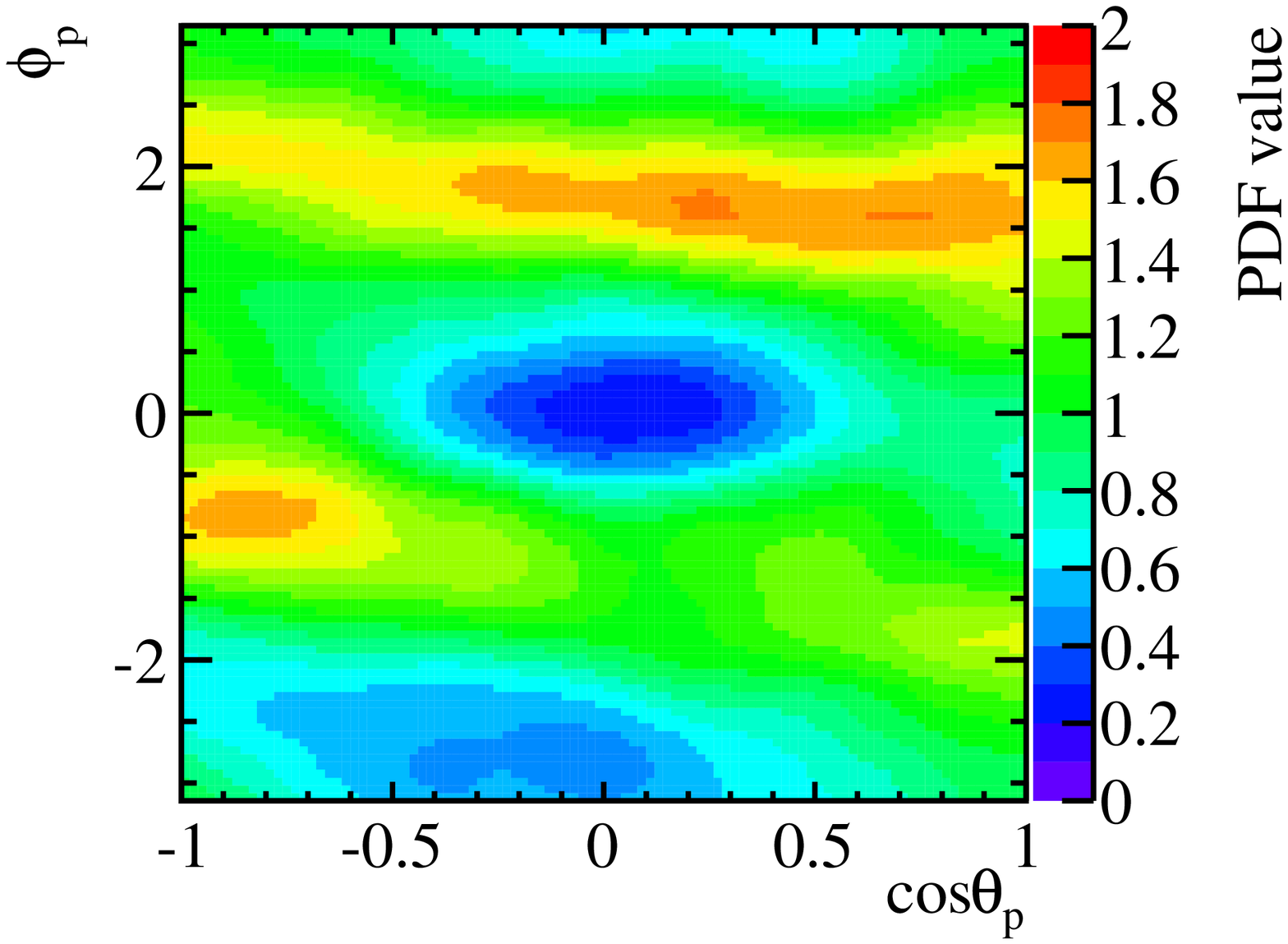}
  \put(-60, 130){(b)}

  \includegraphics[width=0.32\textwidth]{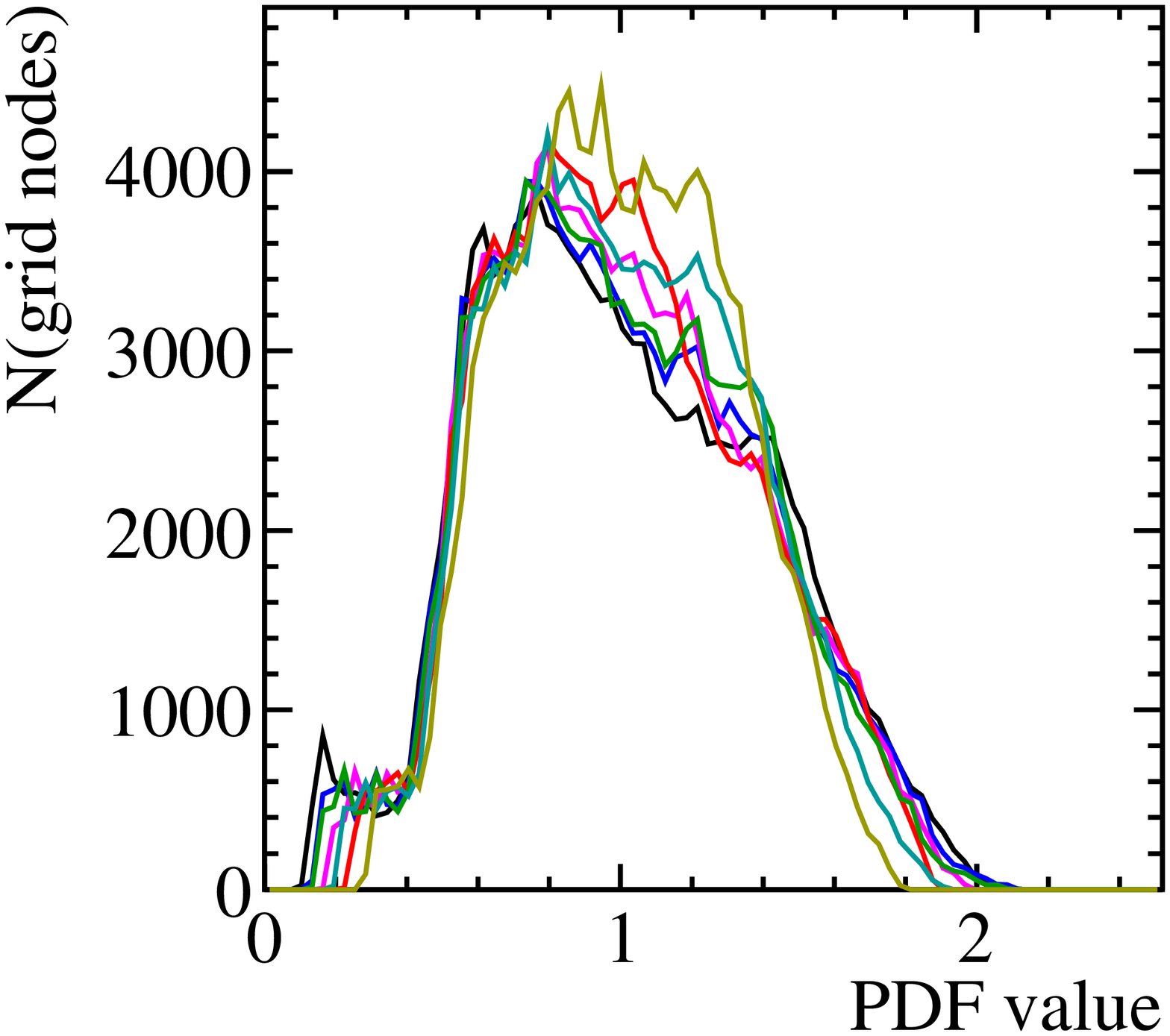}
  \put(-30, 110){(c)}
  \includegraphics[width=0.32\textwidth]{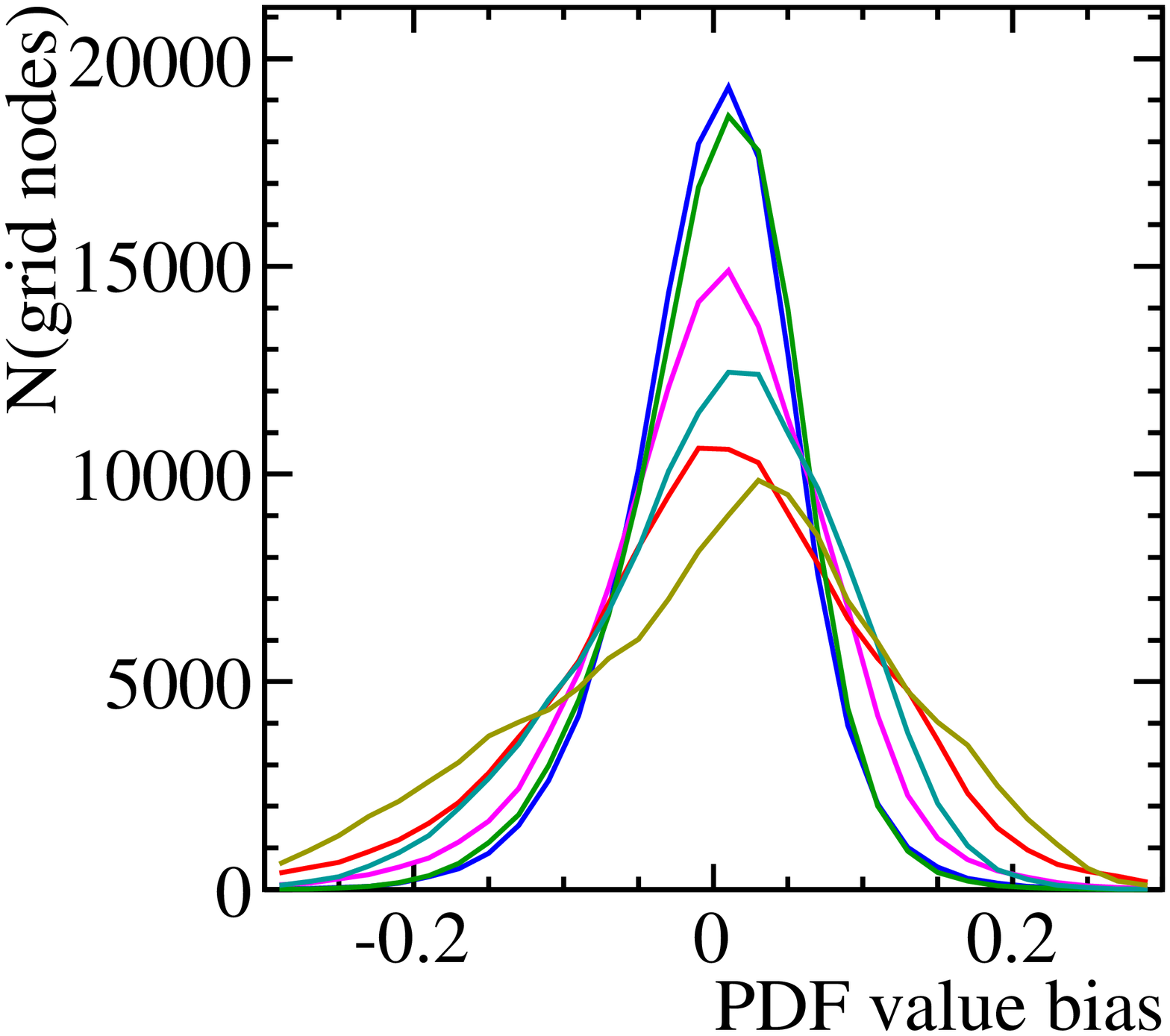}
  \put(-30, 110){(d)}
  \includegraphics[width=0.32\textwidth]{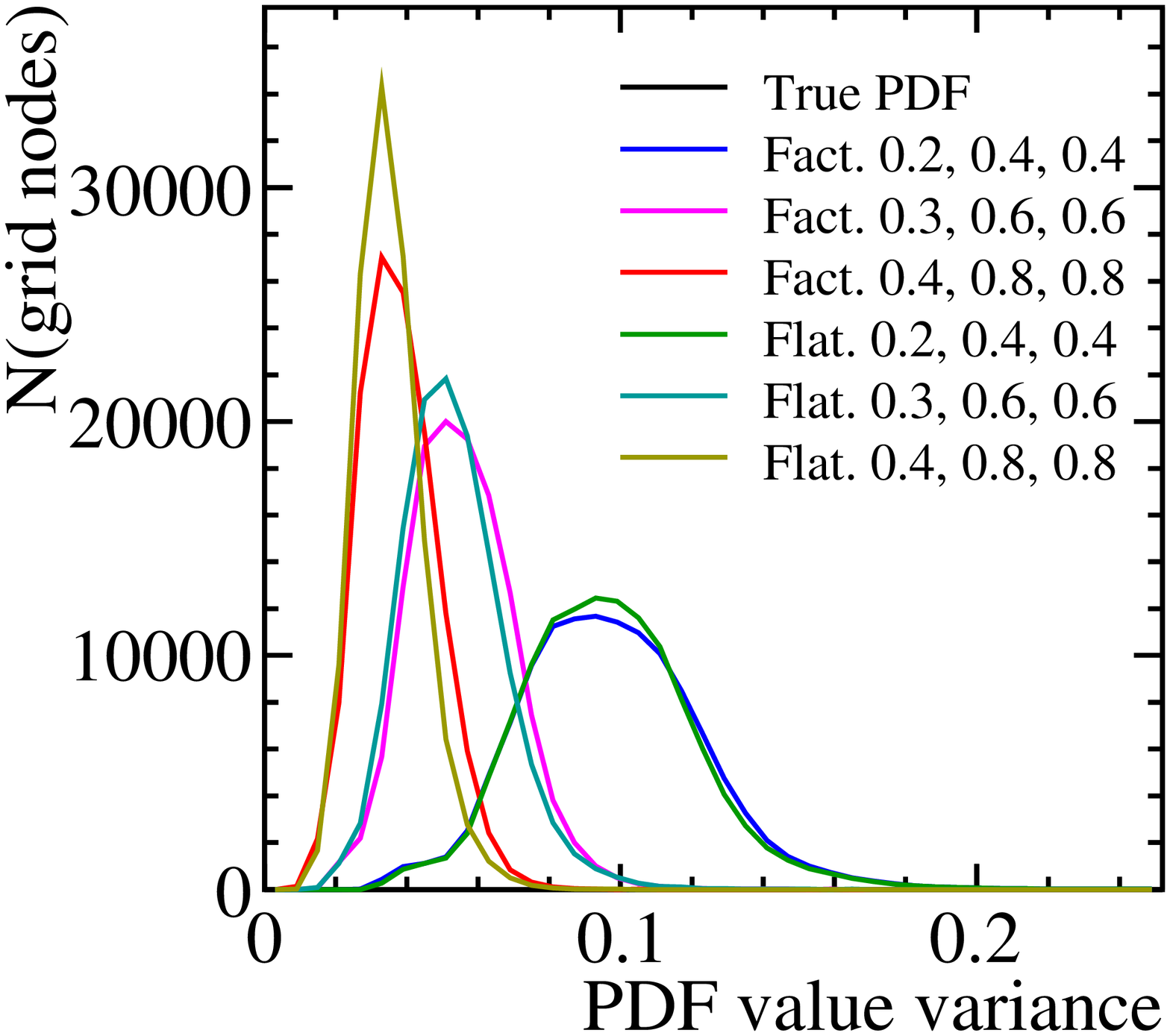}
  \put(-30, 40){(e)}
  \caption{Results of the estimation of the efficiency of $\lbdnppi$ decays in angular variables. (a) Slice of reference 
           density obtained with narrow kernel and (b) an example of the same slice obtained with $10^5$
           sample size and kernel widths of ($0.3$, $0.6$, $0.6$). (c) Distribution of PDF values, 
           and distributions of (d) average bias and (e) variance of estimated PDF values. }
  \label{fig:angular}
\end{figure}

As above, 
we use the large sample with narrow kernel to obtain the reference PDF. Since the phase space 
is now three-dimensional, we can only plot its slice in a pair of variables. As an example, the slice in 
$(\cos\theta_p, \phi_p)$ plane for $\phi_{D\pi}=\pi/4$ is shown in Fig.~\ref{fig:angular} (a). 
Figure~\ref{fig:angular} (b) shows the same slice for a sample of $10^5$ events and the kernel 
widths of $(0.3, 0.6$ rad, $0.6$ rad$)$ in the three variables.  

Instead of presenting the bias and variance of the PDF values as a function of phase space position, 
as we did for the Dalitz plot distribution, we now show the distributions of these values over the points 
across the phase space in the form of histograms. Figures~\ref{fig:angular} (c), (d) and (e) show 
histograms of the PDF value, its bias with respect to the reference density, and variance, respectively, 
where each entry corresponds to a node of the rectangular grid in the three-dimensional angular phase space. 
The value of $\Delta$ is thus the RMS of the distribution in Fig.~\ref{fig:angular} (d), and the 
value of $\sigma$ is the mean of the distribution in Fig.~\ref{fig:angular} (e). 

We test two different options for the approximation PDF used to estimate the angular density. We compare results 
with the uniform approximation PDF (``flat'' approximation), 
and an approximation PDF which is the product of distributions in each 
of the angles (``factorised'' approximation). For the latter, the density for each individual variable is 
in turn the result of applying the 
relative KDE method with the uniform approximation PDF. Results for these two approaches with different kernel 
widths are summarised in Table~\ref{tab:angular}. The factorised approximation PDF gives a 
better quality. 

\begin{table}
  \caption{Results of the estimation of the angular efficiency of \lbdnppi decays. 
           Systematic ($\Delta$), variance ($\sigma$) terms and the overall figure 
           of merit $Q$ for different kernel widths and flat and factorised 
           approximation PDFs. }
  \label{tab:angular}
  \begin{center}
  \begin{tabular}{ll|ccc}
    Kernel width          & Appr. PDF & Bias $\Delta$ & Variance $\sigma$ & $Q$ \\
    \hline
    0.2, 0.4, 0.4        & Flat    & 0.059   & 0.097      & 0.113 \\
    0.3, 0.6, 0.6        & Flat    & 0.085   & 0.053      & 0.100 \\ 
    0.4, 0.8, 0.8        & Flat    & 0.117   & 0.035      & 0.122 \\ 
    0.2, 0.4, 0.4        & Factorised & 0.058   & 0.097      & 0.113 \\
    0.3, 0.6, 0.6        & Factorised & 0.076   & 0.055      & 0.094 \\ 
    0.4, 0.8, 0.8        & Factorised & 0.106   & 0.038      & 0.113 \\ 
  \end{tabular}
  \end{center}
\end{table}

\subsection{Estimation of the full 5D density}

The full 5D density of the decay \lbdnppi is estimated with several different approaches. First, we test the 
factorised PDF which is the product of Dalitz and angular densities. We also try 5D relative KDE 
estimators with different approximation PDFs: uniform (``flat''), factorised PDF made of Dalitz and angular densities (``2D$\times$3D''), as well as factorised PDF which is a product of Dalitz density and densities of each of
the three angular variables (``2D$\times($1D$)^3$''). The kernel widths for the components of the 
factorised PDFs are taken to be those that optimise the $Q$ value in Sections.~\ref{sec:dalitz} and \ref{sec:angular}. 
For the full 5D kernel, we use the width which is larger by a factor $s$ than the kernel width of the component densities. 

\begin{figure}
  \includegraphics[width=0.47\textwidth]{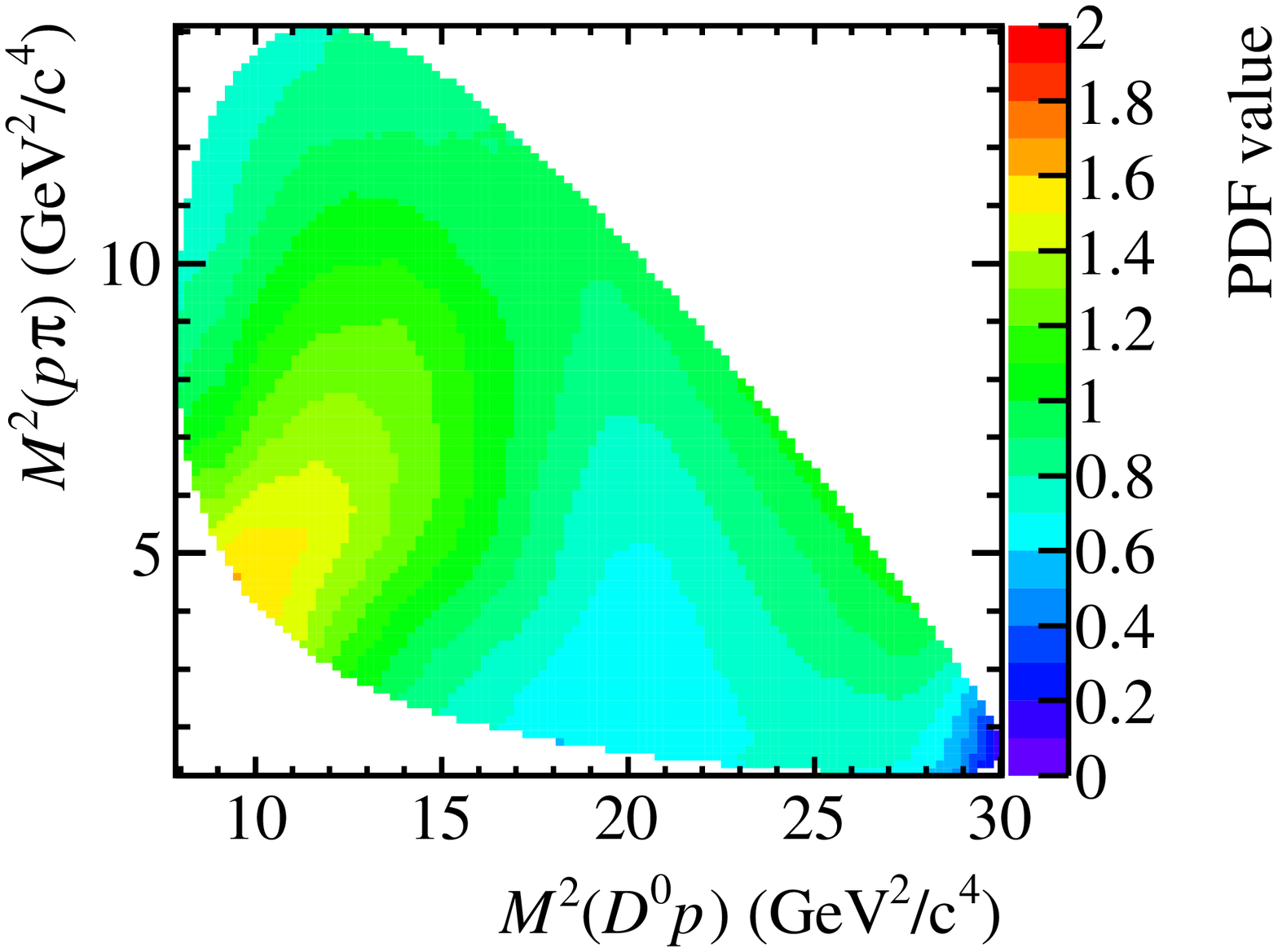}
  \put(-65, 130){(a)}
  \includegraphics[width=0.47\textwidth]{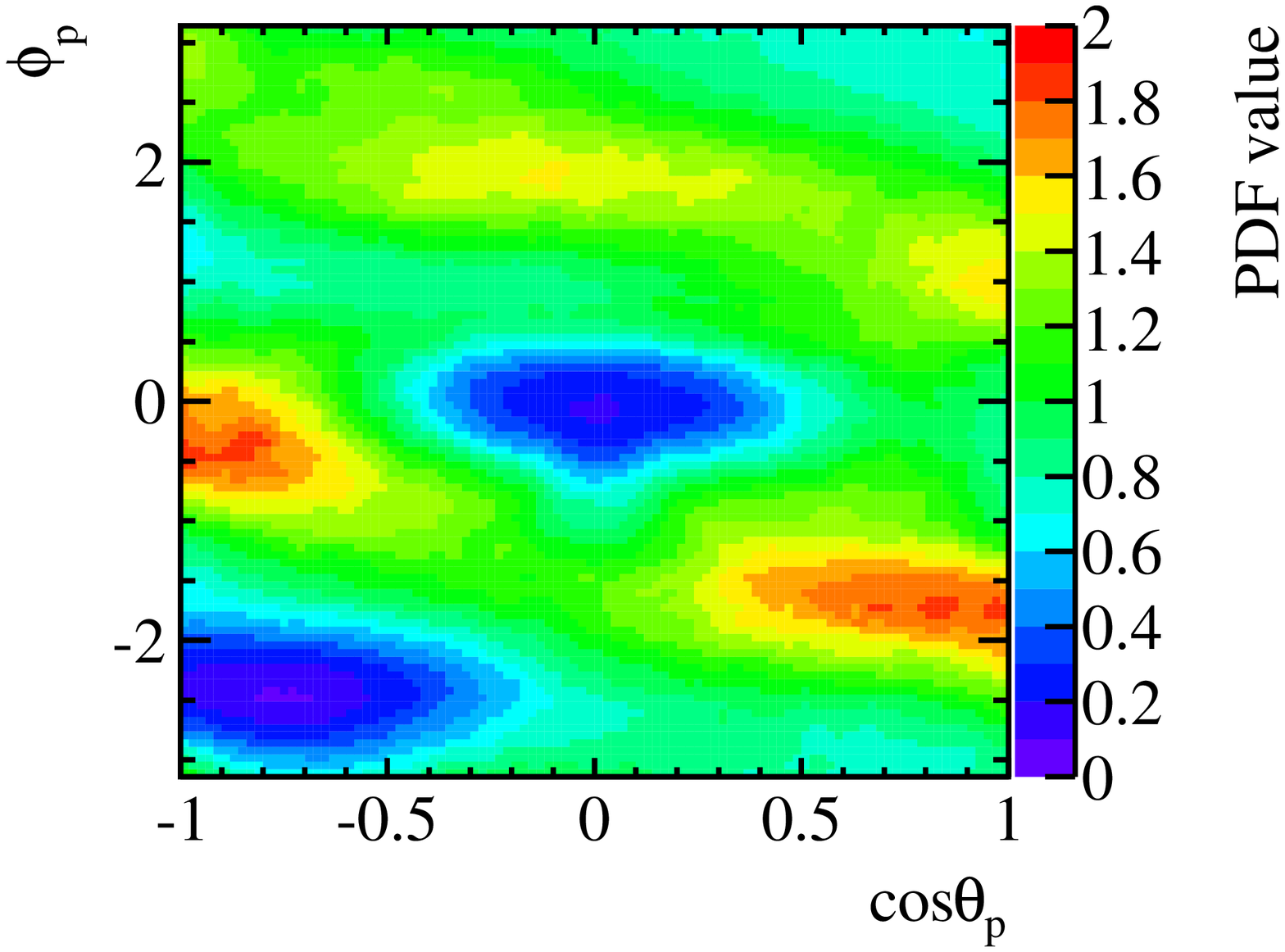}
  \put(-65, 130){(b)}

  \includegraphics[width=0.47\textwidth]{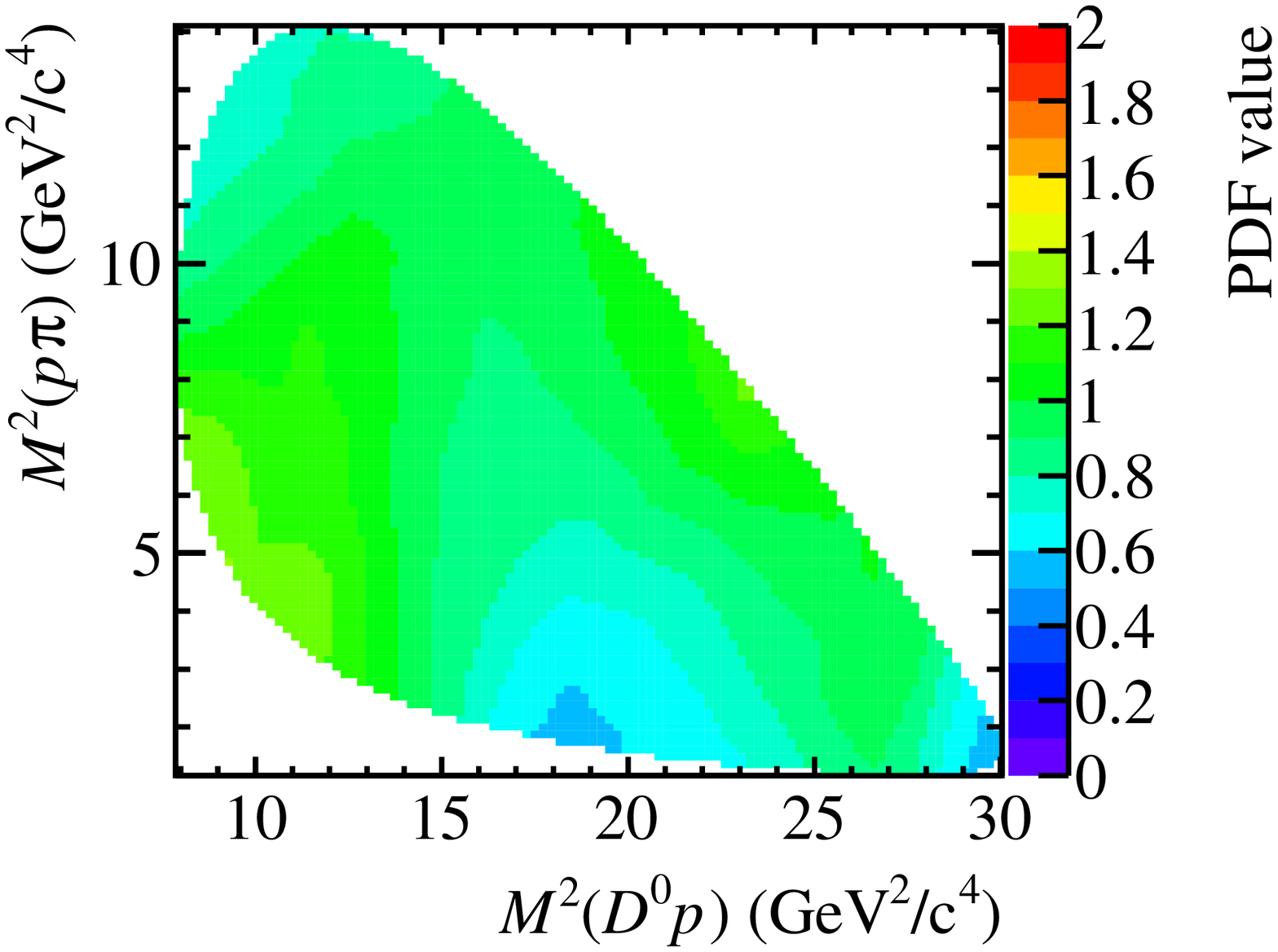}
  \put(-65, 130){(c)}
  \includegraphics[width=0.47\textwidth]{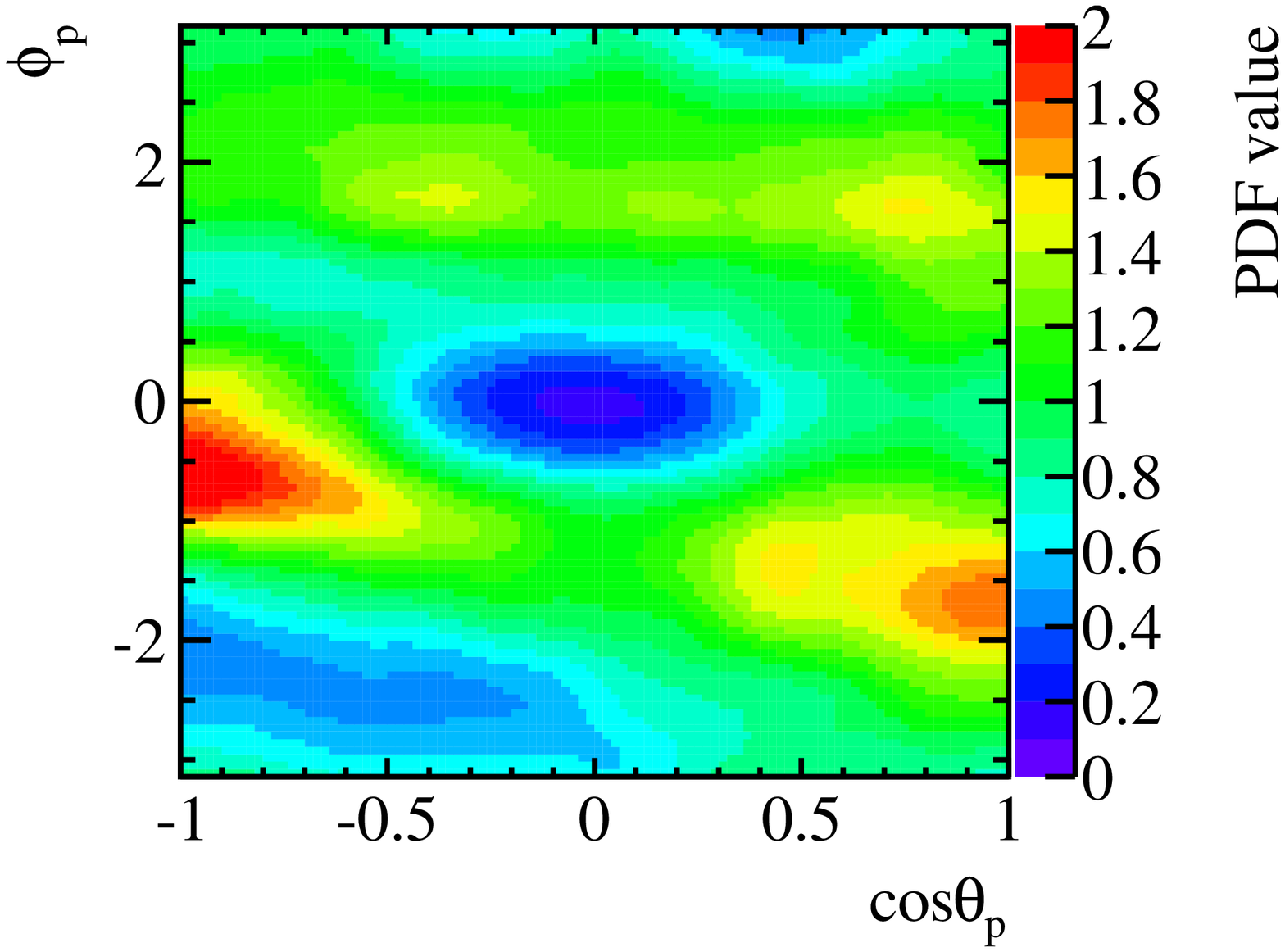}
  \put(-65, 130){(d)}

  \includegraphics[width=0.32\textwidth]{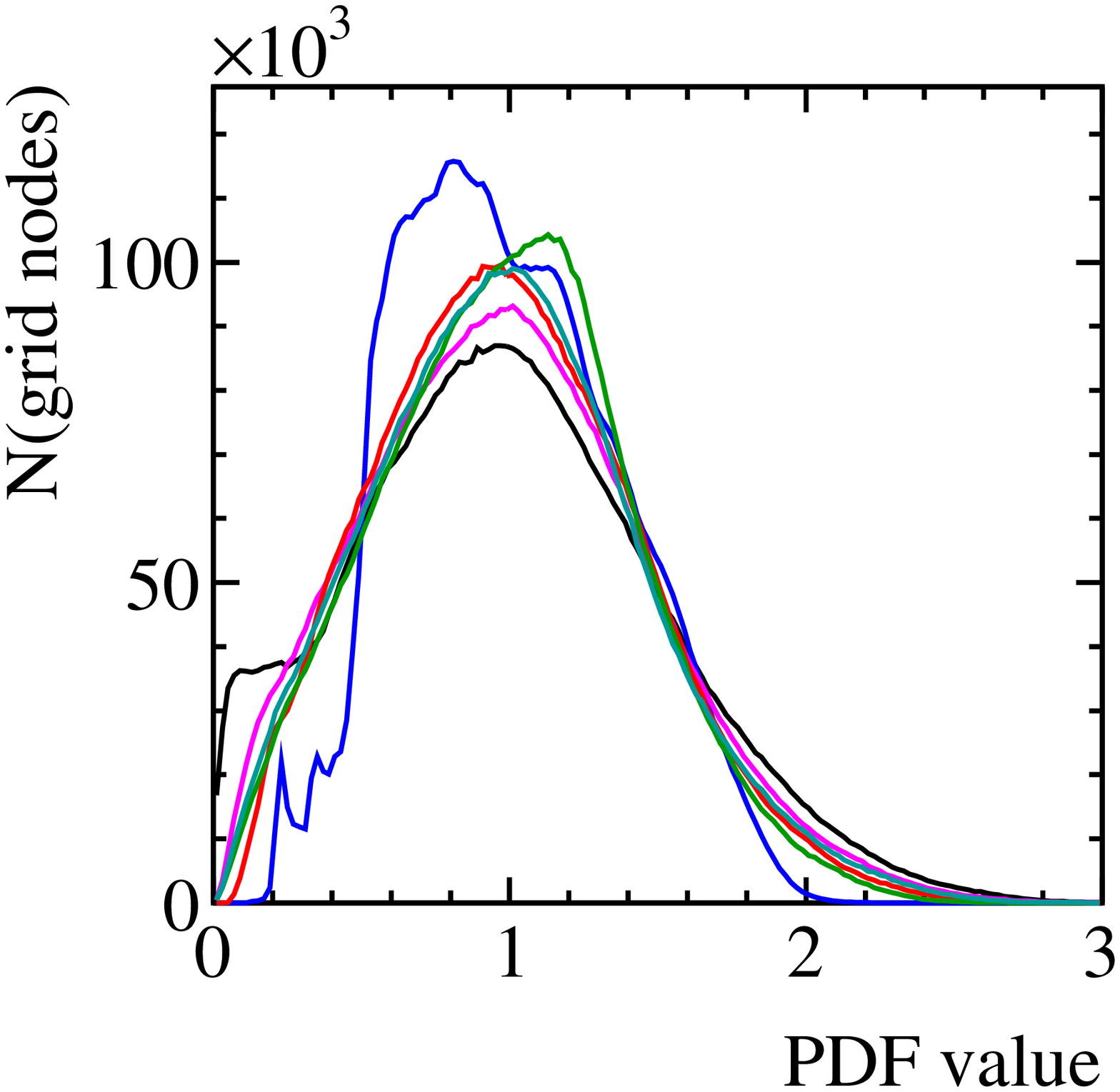}
  \put(-35, 105){(e)}
  \includegraphics[width=0.32\textwidth]{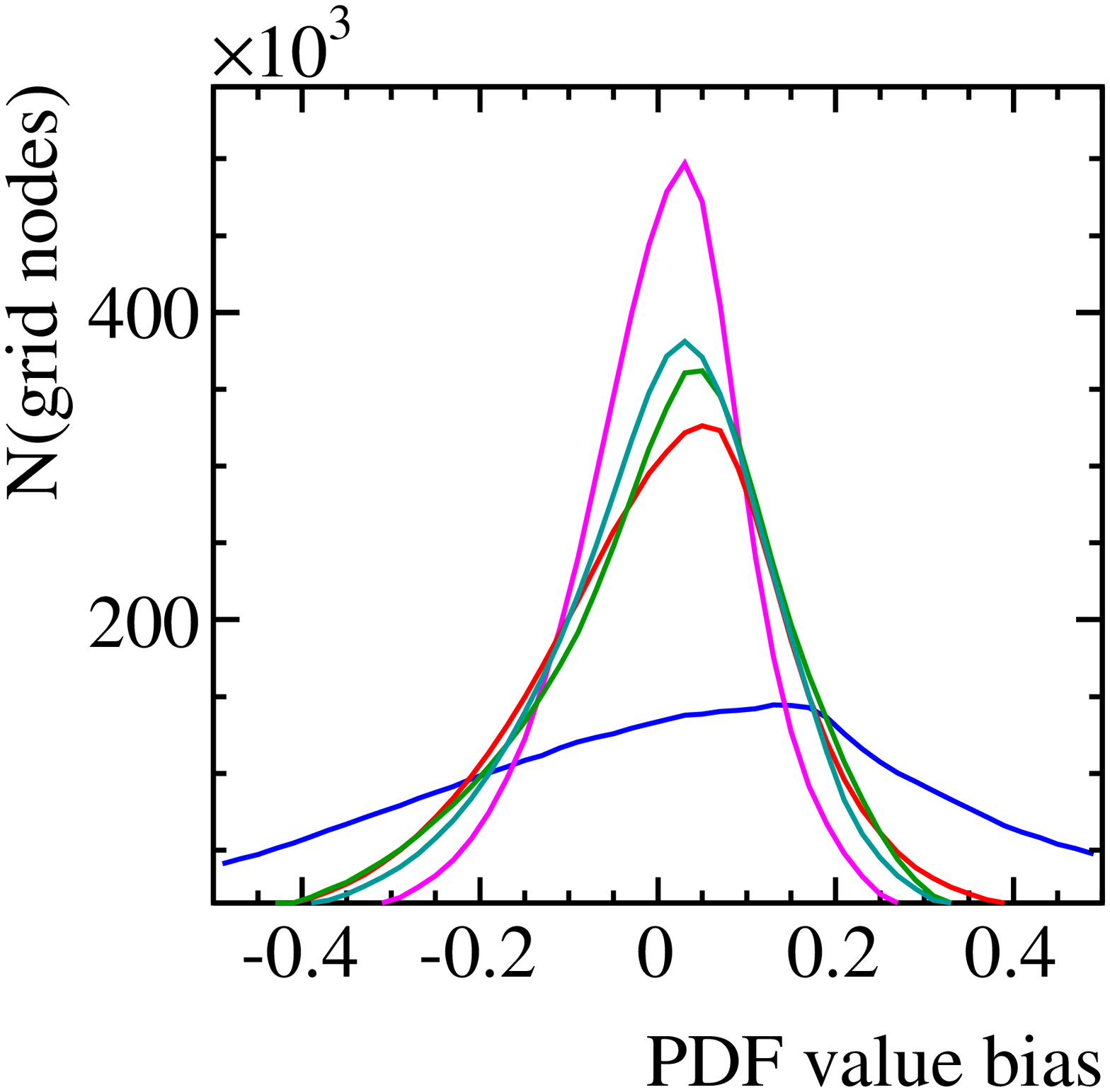}
  \put(-35, 105){(f)}
  \includegraphics[width=0.32\textwidth]{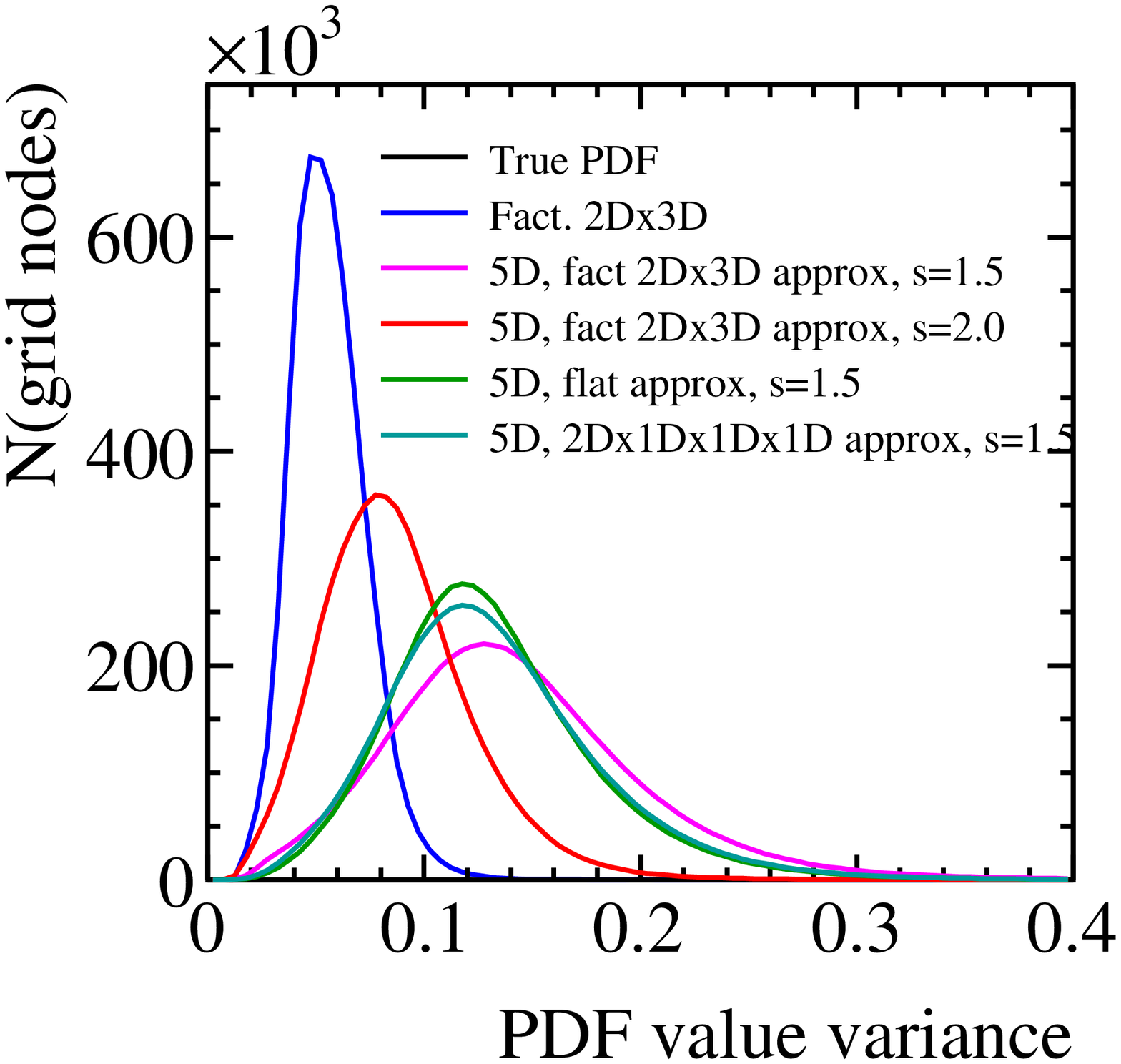}
  \put(-35, 40){(g)}
  \caption{Results of the estimation of the 5D efficiency of $\lbdnppi$ decays. Slice of reference 
           density obtained with narrow kernel in (a) Dalitz plot and (b) angular variables, 
           and (c,d) an example of the same slices obtained with $10^5$
           sample size and kernel widths of (1.5 GeV$^2/c^4$, 1.5 GeV$^2/c^4$, $0.25$, $0.5$, $0.5$). 
           (e) Distribution of PDF values, 
           and distributions of (f) average bias and (g) variance of estimated PDF values. }
  \label{fig:5d}
\end{figure}

The results of the 5D PDF estimation are given in Fig.~\ref{fig:5d}. Figure~\ref{fig:5d} (a) and (b) show 
the Dalitz plot and $(\cos\theta_p, \phi_p)$ slices of the reference PDF which is obtained as explained above. 
The same slices for the PDF estimated using $10^5$ sample with the relative KDE method and 2D$\times$3D approximation 
PDF are shown in Fig.~\ref{fig:5d} (c) and (d). This estimation uses a width factor $s=2$ for the 5D kernel. 
Figures~\ref{fig:5d} (e), (f), and (g) show the distributions of PDF value, PDF bias and its variance, respectively, 
as in Section~\ref{sec:angular}. 
 
The quantitative results of the estimation are given in Table~\ref{tab:5d}. As can be seen from the table, 
the best approach using relative KDE technique (which uses 2D$\times$3D approximation PDF) gives on 
average twice lower figure of merit ($Q=0.178$) of PDF estimation than the product of Dalitz and angular densities ($Q=0.353$). 

\begin{table}
  \caption{Results of the estimation of the 5D efficiency of \lbdnppi decays. 
           Systematic ($\Delta$), variance ($\sigma$) terms and the overall figure 
           of merit $Q$ for different approaches. }
  \label{tab:5d}
  \begin{center}
  \begin{tabular}{ll|ccc}
    Kernel width       & PDF & Bias $\Delta$ & Variance $\sigma$& $Q$ \\
    factor $s$  &     &      &        &         \\
    \hline
    -           & 2D$\times$3D factorised              & 0.349   & 0.056      & 0.353 \\
    2.0         & 5D, fact. 2D$\times$3D approx.       & 0.159   & 0.087      & 0.181 \\ 
    1.5         & 5D, fact. 2D$\times$3D approx.       & 0.108   & 0.142      & 0.178 \\
    1.5         & 5D, flat approximation               & 0.150   & 0.132      & 0.200 \\
    1.5         & 5D, fact. 2D$\times($1D$)^3$ approx. & 0.140   & 0.132      & 0.192 \\
  \end{tabular}
  \end{center}
\end{table}

\section{Conclusion}

A simple modification of the kernel density estimation technique is proposed to 
correct for boundary effects and other expected narrow (with respect to the kernel width) structures. 
In this approach, the estimated PDF is represented as the product of the approximated PDF, 
which describes the behaviour near the boundaries and narrow structures, and the 
kernel density PDF. 

This approach is shown to work well in the case of description of the five-dimensional efficiency shape 
for the decay \lbdnppi. Approximation PDF is the 
factorised PDF of the phase space variables, while the relative KDE provides a 
smooth correction to take into account the correlations between the phase space variables. 
This allows to use a relatively wide kernel for the multidimensional PDF estimation which 
reduces fluctuations due to limited sample size, while still keeping the essential features 
of the PDF ({\it e.g.} near the boundaries of the phase space) not smeared by the kernel. 

\acknowledgments

The author is grateful to his colleagues from the LHCb collaboration for useful discussions about the 
method, feedback on the usage of {\tt Meerkat} package, and suggestions to improve the text of the paper: 
Thomas Blake, Tim Gershon, Michal Kreps, Thomas Latham, Mark Whitehead (University of Warwick), 
Michael Williams (MIT), Peter Griffith (University of Birmingham). 

This work is supported by the Science and Technology Facilities Council (United Kingdom),
by the Russian Foundation for Basic Research grant 14-02-00569 A, and 
by the Russian Ministry of Education and Science contract 14.610.21.0002.

\end{document}